\documentclass[preprint,aps,superscriptaddress,nofootinbib,tightenlines,floatfix]{revtex4}

\usepackage{graphicx}
\usepackage{bm}
\usepackage{comment}
\usepackage{color}

\def\si{^1 \hskip -0.03in S _0}
\def\siii{^3 \hskip -0.025in S _1}
\def\diii{^3 \hskip -0.045in D _1}

\def\latta{24^3\times 48}
\def\lattb{32^3\times 48}

\def\b{0.145}
\def\bfm{\b~{\rm fm}}

\def\blatt{0.1453(16)~{\rm fm}}
\def\La{3.4}
\def\Lb{4.5}
\def\Lc{6.7}
\def\Lafm{\La~{\rm fm}}
\def\Lbfm{\Lb~{\rm fm}}
\def\Lcfm{\Lc~{\rm fm}}
\def\Ta{6.7}
\def\Tb{6.7}
\def\Tc{9.0}
\def\NcfgA{3822} 
\def\NscrA{96}
\def\NcfgB{3050} 
\def\NscrB{72}
\def\NcfgC{1905} 
\def\NscrC{54}

\def\mpilu{0.59426(12)(11)~{\rm l.u.} }
\def\mBlu{1.20359(41)(61)~{\rm l.u.} }

\def\mpi{ 800 }
\def\mpiMeV{ \mpi~{\rm MeV} }
\def\mpifull{ 805.9(0.6)(0.4)(8.9) }
\def\mpifullMeV{ \mpifull~{\rm MeV} }

\def\mBfull{ 1.635(0)(0)(18)~{\rm GeV} }

\def\Bd{19.5(3.6)(3.1)(0.2)  }
\def\Bnn{15.9(2.7)(2.7)(0.2) }
\def\BdMeV{ \Bd~{\rm MeV}  }
\def\BnnMeV{ \Bnn~{\rm MeV} }

\def\ampising{9.50^{+0.78}_{-0.69}{}^{+1.10}_{-0.80}}
\def\rmpising{4.61^{+0.29}_{-0.31}{}^{+0.24}_{-0.26}}
\def\ampisingphys{2.33^{+0.19}_{-0.17}{}^{+0.27}_{-0.20}}
\def\rmpisingphys{1.130^{+0.071}_{-0.077}{}^{+0.059}_{-0.063}}

\def\ampitrip{7.45^{+0.57}_{-0.53}{}^{+0.71}_{-0.49}}
\def\rmpitrip{3.71^{+0.28}_{-0.31}{}^{+0.28}_{-0.35}}
\def\ampitripphys{1.82^{+0.14}_{-0.13}{}^{+0.17}_{-0.12}}
\def\rmpitripphys{0.906^{+0.068}_{-0.075}{}^{+0.068}_{-0.084}}

\newcount\hour \newcount\hourminute \newcount\minute 
\hour=\time \divide \hour by 60
\hourminute=\hour \multiply \hourminute by 60
\minute=\time \advance \minute by -\hourminute
\newcommand{\mydate}{\ \today \ - \number\hour :\number\minute}

\begin{document}

\begin{figure}[!t]

  \vskip -1.5cm
  \leftline{\includegraphics[width=0.25\textwidth]{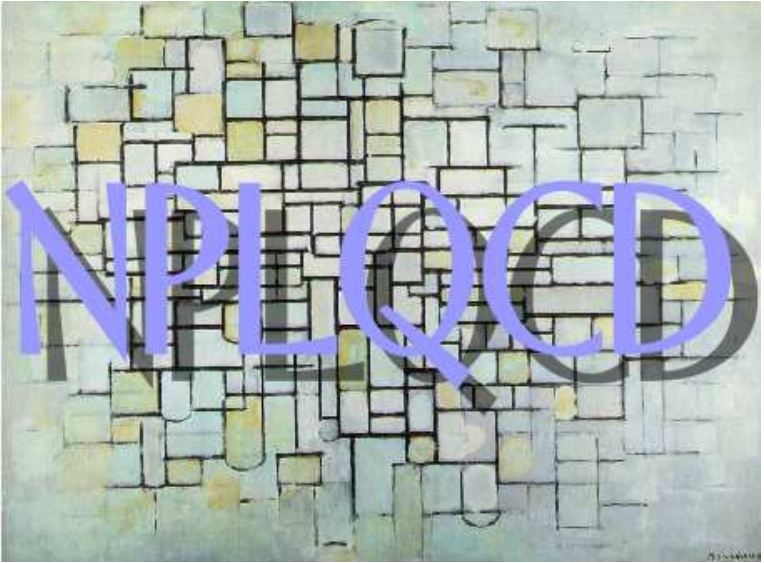}}
\end{figure}

\preprint{
\vbox{ 
\hbox{MIT-CTP-4434}
\hbox{ICCUB-13-040}
\hbox{JLAB-THY-13-1689}
\hbox{NT@UW-13-05}
\hbox{NT-LBNL-13-004}
\hbox{UCB-NPAT-13-004}
\hbox{UNH-13-2}
}}

\title{Nucleon-Nucleon Scattering Parameters in
  the Limit of\\ SU(3) Flavor Symmetry }

\author{S.R.~Beane} \affiliation{Helmholtz-Institut f\"ur Strahlen- und Kernphysik (Theorie),
Universit\"at Bonn, D-53115 Bonn, Germany}

\author{E.~Chang} \affiliation{Department of Physics, University of
  Washington, Box 351560, Seattle, WA 98195, USA}

\author{S.D.~Cohen} \affiliation{Department of Physics, University of
  Washington, Box 351560, Seattle, WA 98195, USA}

\author{W.~Detmold} \affiliation{
Center for Theoretical Physics, Massachusetts Institute of Technology, Cambridge, MA 02139, USA}

\author{P. Junnarkar} \affiliation{Department of Physics, University of
  New Hampshire, Durham, NH 03824-3568, USA}

\author{H.W.~Lin} \affiliation{Department of Physics, University of
  Washington, Box 351560, Seattle, WA 98195, USA}

\author{T.C.~Luu} \affiliation{N Section, Lawrence Livermore National
  Laboratory, Livermore, CA 94551, USA}

\author{K.~Orginos} \affiliation{Department of Physics, College of
  William and Mary, Williamsburg, VA 23187-8795, USA}
\affiliation{Jefferson Laboratory, 12000 Jefferson Avenue, Newport
  News, VA 23606, USA}

\author{A.~Parre\~no} \affiliation{Dept. d'Estructura i Constituents
  de la Mat\`eria.  Institut de Ci\`encies del Cosmos (ICC),
  Universitat de Barcelona, Mart\'{\i} i Franqu\`es 1, E08028-Spain}

\author{M.J.~Savage} 
\affiliation{Helmholtz-Institut f\"ur Strahlen- und Kernphysik (Theorie),
Universit\"at Bonn, D-53115 Bonn, Germany}\affiliation{Department of Physics, University of
  Washington, Box 351560, Seattle, WA 98195, USA}

\author{A.~Walker-Loud} \affiliation{Lawrence Berkeley National
  Laboratory, Berkeley, CA 94720, USA}\affiliation{Department of Physics, 
University of California, Berkeley, CA 94720, USA}

\collaboration{ NPLQCD Collaboration }

\date{\mydate}

\begin{abstract}
  \noindent
  The scattering lengths and effective ranges that describe low-energy
  nucleon-nucleon scattering are calculated in the limit of
  SU(3)-flavor symmetry at the physical strange-quark mass with
  Lattice Quantum Chromodynamics.  The calculations are performed with
  an isotropic clover discretization of the quark action in three
  volumes with spatial extents of $L\sim\Lafm, \Lbfm$ and $\Lcfm$, and
  with a lattice spacing of $b\sim \bfm$.  With determinations of the
  energies of the two-nucleon systems (both of which contain bound
  states at these up and down quark masses) at rest and moving in the
  lattice volume, L\"uscher's method is used to determine the
  low-energy phase shifts in each channel, from which the scattering
  length and effective range are obtained. The scattering parameters,
  in the $\si$ channel are found to be $m_\pi a^{(\si)} = \ampising$ and
  $m_\pi r^{(\si)} = \rmpising$, and in the $\siii$ channel are $m_\pi
  a^{(\siii)}=\ampitrip$ and $m_\pi r^{(\siii)} = \rmpitrip$. These
  values are consistent with the two-nucleon system exhibiting
  Wigner's supermultiplet symmetry, which becomes exact in the limit
  of large-$N_c$. In both spin channels, the phase shifts 
change sign at higher momentum, near the start of the t-channel cut,
indicating that the nuclear interactions have a repulsive core even
at the SU(3)-symmetric point.
\end{abstract}
\maketitle
\vfill\eject
%

\section{Introduction}
\label{sec:intro}
\noindent
Decades of experimental measurements of the nucleon-nucleon (NN) cross
sections have resulted in precise phase-shifts and scattering
parameters that provide the cornerstone of nuclear physics.  These
two-body interactions, when combined with multi-body interactions,
dictate the low-energy spectrum and interactions of nuclei, and also
the equation of state of nuclear matter at moderate densities.
Further, these interactions are responsible for the fine-tunings that
permeate nuclear physics, and are responsible for producing sufficient
carbon in the universe to allow for the emergence of life.  During the
last several years, there has been a substantial effort to determine
the NN interactions directly from quantum chromodynamics
(QCD) using the numerical technique of Lattice QCD
(LQCD)~\cite{Fukugita:1994na,Fukugita:1994ve,Beane:2006mx,Beane:2009py,Ishii:2006ec,Aoki:2008hh,Aoki:2009ji,
  Yamazaki:2011nd,Yamazaki:2009ua,deForcrand:2009dh,Beane:2011iw,Inoue:2011ai,Beane:2012vq,Yamazaki:2012hi}.
Steady progress is being made toward this objective, but calculations
at the physical light-quark masses have not yet been performed, and
essentially only one lattice spacing has been used in calculations.
While calculations at the physical light-quark masses, that are
extrapolated to the continuum limits and are performed in volumes (and
temporal directions) that are much larger than the inverse Compton
wavelength of the pion, are required to verify the LQCD technology and
provide a rigorous underpinning of modern nuclear interactions,
calculations at heavier pion masses are equally important in
understanding and quantifying the fine-tunings in nuclear physics.  It
is crucial to understand how the fine-tunings in nuclear physics
translate into constraints on the five relevant fundamental parameters
in the standard model of particle physics: the three light-quark
masses, and the strong and electromagnetic coupling constants, and the
NN interaction provides the simplest place to begin this
investigation.

One reason that there are presently few LQCD calculations of
NN interactions is the significantly greater complexity
of multinucleon systems as compared with systems of single mesons and
baryons.  A second reason is that significant computational resources
are required to generate high-quality ensembles of gauge field
configurations at or near the physical light-quark masses, an effort
that has only become practical with the availability of petascale
computers, and, as yet, these ensembles are not at sufficiently large
volume to be of use in nuclear physics. At heavier quark masses, the
resources required to generate ensembles of lattice gauge
configurations and light-quark propagators are relatively small, the
degradation of the signal-to-noise in multinucleon correlation
functions is significantly reduced, and thermal effects are
exponentially suppressed, compared to calculations at lighter pion
masses.  For these reasons, we performed LQCD calculations of a number
of s-shell nuclei and hypernuclei with $A\le 5$ at the SU(3)-flavor
symmetry point with the physical strange quark mass giving $m_\pi =
\mpifullMeV$, at a lattice spacing of $b \sim \bfm$, and in lattice
volumes with spatial extents $L\sim\Lafm, \Lbfm$ and
$\Lcfm$~\cite{Beane:2012vq}.  In this work, we continue this study and
explore the NN scattering phase shifts below the
inelastic threshold and the associated scattering parameters relevant
below the t-channel cut at the SU(3) symmetric point.

\section{Overview of the Lattice QCD Calculations}
\label{sec:LQCD}
\noindent
Three ensembles of isotropic gauge-field configurations, generated
with a tadpole-improved L\"uscher-Weisz gauge action and a clover
fermion action~\cite{Sheikholeslami:1985ij}, are used in this work and
have been used previously to calculate the lowest-lying levels of the s-shell
nuclei and hypernuclei~\cite{Beane:2012vq}.  This particular
lattice-action setup follows closely the anisotropic clover action of
the ensembles generated by the JLab group that we have used in our
previous
calculations~\cite{Beane:2009py,Beane:2009gs,Beane:2010hg,Beane:2011iw,Beane:2011xf,Beane:2012ey}.
The parameter tuning and scaling properties of this action will be
discussed elsewhere~\cite{ISO}.  One level of stout
smearing~\cite{Morningstar:2003gk} with $\rho=0.125$ and
tadpole-improved tree-level clover coefficient $c_{\rm SW}=1.2493$ are
used in the gauge-field generation.
Studies~\cite{Hoffmann:2007nm,Edwards:2008ja,ISO} of the 
partially-conserved axial-current (PCAC) relation
in the Schr\"odinger functional indicate that this choice is
consistent with vanishing ${\cal O}(b)$ violations, leading to
discretization effects that are essentially ${\cal O}(b^2)$.  The
parameters of the ensembles are listed in
Table~\ref{tab:gauageparams}, and further details will be presented
elsewhere~\cite{ISO}.  As two-nucleon systems are the focus of this
work, relatively large lattice volumes are employed for the
calculations, with correspondingly large values of $m_\pi L$ and
$m_\pi T$.  In order to convert the calculated energies from
lattice units (l.u.)  into physical units (MeV), a lattice spacing of
$b=\blatt$ has been determined for these ensembles of gauge-field
configurations from the $\Upsilon$ spectrum~\cite{meinelPRIV}.
\begin{table}
\begin{center}
\begin{minipage}[!ht]{16.5 cm}
  \caption{Parameters of the ensembles of gauge-field configurations and of the measurements used in this work.
    The lattices have dimension  $L^3\times T$, a lattice spacing $b$, and
    a bare quark mass $b\ m_q$ (in lattice units) generating a pion of
    mass $m_\pi$. 
$N_{\rm src}$ light-quark sources are used (as described in the text)
to perform measurements on $N_{\rm cfg}$ configurations in each ensemble.
The three uncertainties associated with the pion mass are
statistical, fitting systematic and that associated with the 
lattice spacing, respectively.
  }  
\label{tab:gauageparams}
\end{minipage}
\setlength{\tabcolsep}{0.3em}
\begin{tabular}{c|cccccccccccc}
\hline
      Label & $L/b$ & $T/b$ & $\beta$ & $b\ m_q$ & $b$ [fm]  & $L$ [fm] & $T$
      [fm] & $m_\pi$ [MeV] & $m_\pi L$ & $m_\pi T$ & $N_{\rm cfg}$ & $N_{\rm src}$\\
\hline
      A& 24 & 48 & 6.1 &-0.2450 & \b& \La & \Ta & 806.5(0.3)(0)(8.9) & 14.3 & 28.5 & \NcfgA & \NscrA\\
      B&       32 & 48 & 6.1 &-0.2450 & \b  &  \Lb & \Tb & 806.9(0.3)(0.5)(8.9) & 19.0 & 28.5  & \NcfgB & \NscrB\\	
      C&48 & 64 & 6.1 &-0.2450 & \b  &  \Lc & \Tc & 806.7(0.3)(0)(8.9) & 28.5 & 38.0 & \NcfgC & \NscrC\\
\hline
\end{tabular}
\begin{minipage}[t]{16.5 cm}
\vskip 0.0cm
\noindent
\end{minipage}
\end{center}
\end{table}     

The $N_{\rm cfg}$ gauge configurations in each of the ensembles are
separated by ten hybrid Monte-Carlo (HMC) evolution
trajectories to reduce autocorrelations, and an average of $N_{\rm
  src}$ measurements are performed on each configuration.  The quark
propagators are constructed with gauge-invariant Gaussian-smeared
sources with stout-smeared gauge links.  These sources are distributed
over a grid, the center of which is randomly distributed within the
lattice volume on each configuration, and the quark propagators are
computed using the BiCGstab algorithm with a tolerance of $10^{-12}$
in double precision.  Quark propagators, either unsmeared or smeared
at the sink using the same parameters as used at the source, give rise
to two sets of correlation functions for each combination of source
and sink interpolating fields, labeled as SP and SS, respectively.
The propagators are contracted to form nucleon blocks projected to
fixed momentum at the sink for use in the calculation of the
correlation functions.  The blocks are defined
as
\begin{equation}
  \label{eq:blockdef}
  {\cal B}^{ijk}_{N}({\bf p},t;x_0)= \sum_{\bf x}e^{i{\bf p}\cdot{\bf
      x}} S_{i}^{(f_1),i^\prime} ({\bf x},t;x_0)S_{j}^{(f_2),j^\prime}({\bf x},t;x_0)
  S_{k}^{(f_3),k^\prime}({\bf x},t;x_0) b^{(N)}_{i^\prime j^\prime k^\prime}
\ \ \ ,
\end{equation}
where $S^{(f)}$ is a quark propagator of flavor $f$, and the indices
are combined spin-color indices running over
$i=1,\ldots,N_cN_s$.\footnote{To be specific, for a quark spin
  component $i_s=1,\ldots,N_s$ and color component $i_c=1,\ldots,N_c$,
  the combined index $i= N_c(i_s-1)+i_c$.}  The choice of the $f_i$
and the tensor $b^{(N)}$ depend on the spin and isospin of the nucleon
under consideration. For our calculations we used the local
interpolating fields constructed in Ref.~\cite{Basak:2005ir},
restricted to those that contain only upper spin components (in the
Dirac spinor basis). This choice results in the simplest interpolating
fields that also have the best overlap with the nucleon ground states.
Blocks are constructed for all lattice momenta $|{\bf p}|^2<4$
allowing for the study of multinucleon systems with zero or nonzero
total momentum and with nontrivial spatial wave functions.
Interpolating operators used at the sink have blocks with back-to-back
momentum to access excited states.  These interpolating operators have
small overlaps onto the ground state, and it is found that the
correlation functions from these back-to-back interpolating operators
are as well fit by single states as are the ground-state correlation
functions. More sophisticated methods such as Matrix
Prony~\cite{Beane:2009kya} and GPoF~\cite{Aubin:2010jc,ko} have been
applied to the correlation functions 
and do not provide significant improvement in the extractions, indicating that
the correlators with back-to-back momenta are close to
orthogonal. Future investigations with a fully variational basis of
operators are required to give us confidence in extracting states
beyond the first excitation.  The dispersion relations of the single
mesons and nucleons on these ensembles have been examined in
Ref.~\cite{Beane:2012vq}, and the infinite volume extrapolations of
the masses are $m_\pi=\mpilu=\mpifullMeV$ and $M_N = \mBlu = \mBfull$.

\section{Nucleon-Nucleon Scattering in the $\si$ Channel}
\label{sec:NNscatt}
\noindent
In contrast to the real world, there is a bound state in the
$\si$-channel at the SU(3)-symmetric point with a binding energy of
$B_{nn} = \BnnMeV$~\cite{Beane:2012vq}~\footnote{It should be noted
  that the HALQCD collaboration does not find a bound state in the
  $\si$ channel nor $\siii-\diii$ coupled-channels at this pion
  mass~\cite{Inoue:2011ai}.  Their results are arrived at through the
  solution of the Schr\"odinger equation with ``potentials''
  calculated on the lattice, a method which is theoretically unsound
  in various ways~\cite{Beane:2010em,Birse:2012ph}, and, moreover, is
  considerably less direct than determining binding energies using
  simple spectroscopy.}.  Two-nucleon correlation functions with total
momentum $|{\rm P}|=0$~\footnote{Here, $|{\rm P}|$ denotes the
  magnitude of the total momentum in units of $2\pi/L$.} and $|{\rm
  P}|=1$ evaluated in three lattice volumes, with spatial extent
$L=\Lafm, \Lbfm, \Lcfm$, are used to extract the binding energy of the
di-neutron.  While the results are found to be consistent, volume
effects are observed in the smallest volume and the binding energy
determined in the largest volume is taken as the infinite volume
value.  In the largest volume, the exponentially suppressed deviations
from the infinite volume value are negligible for this binding energy.
With one bound state in this channel, Levinson's theorem dictates that
the phase shift is $\delta^{(\si)}(0) = \pi$ at threshold.

\begin{figure}[!ht]
  \centering
  \includegraphics[width=0.32\textwidth]{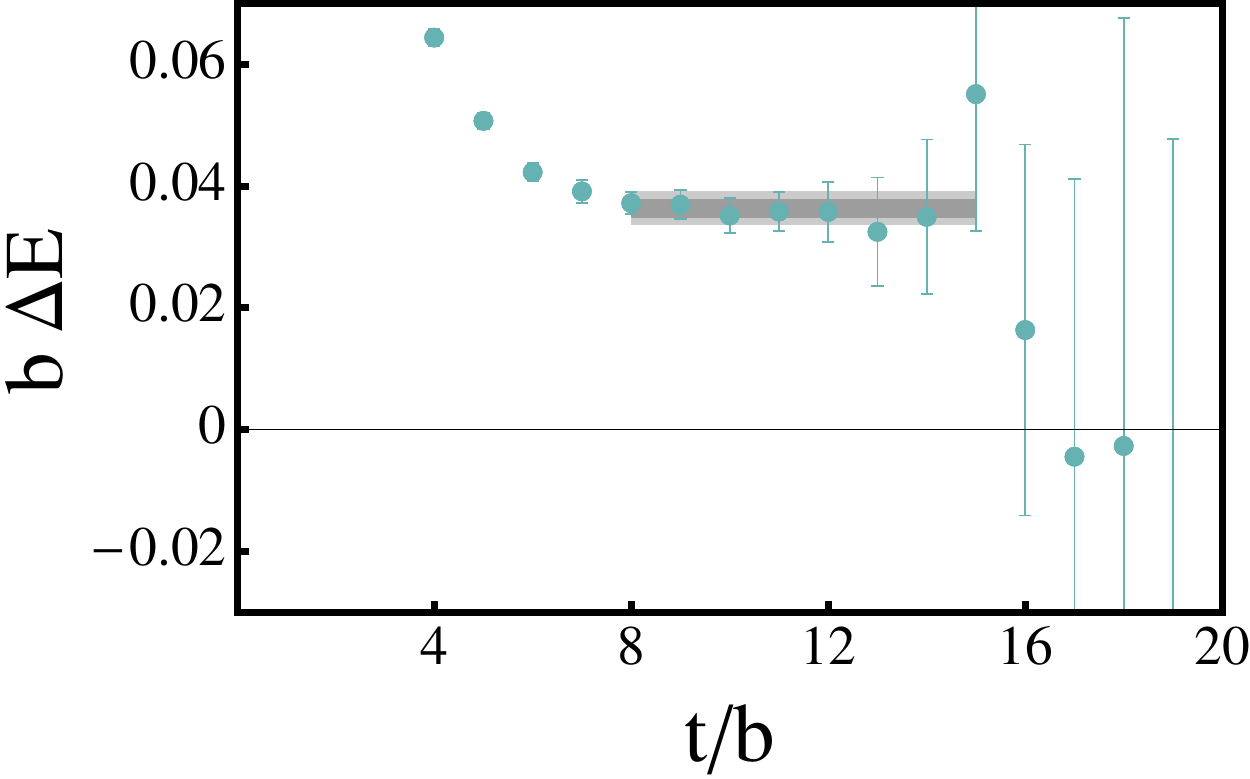}\,
  \includegraphics[width=0.32\textwidth]{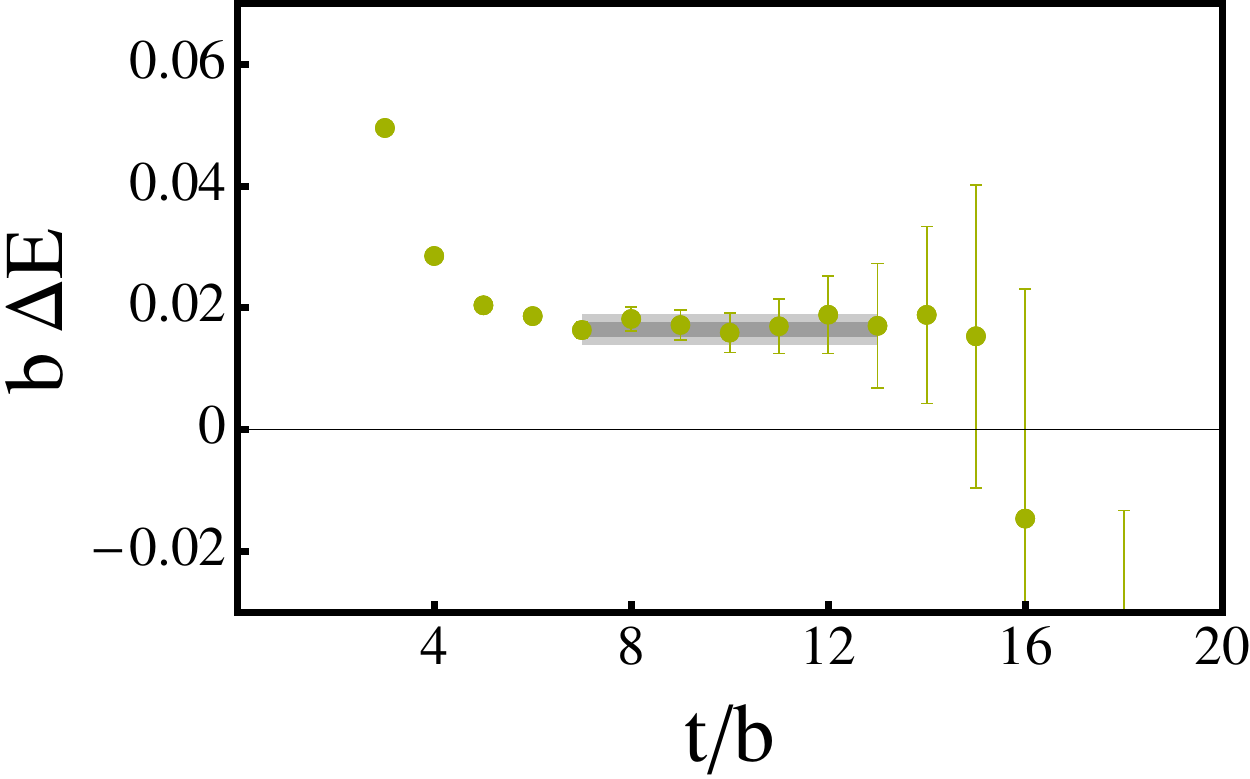} \,
  \includegraphics[width=0.32\textwidth]{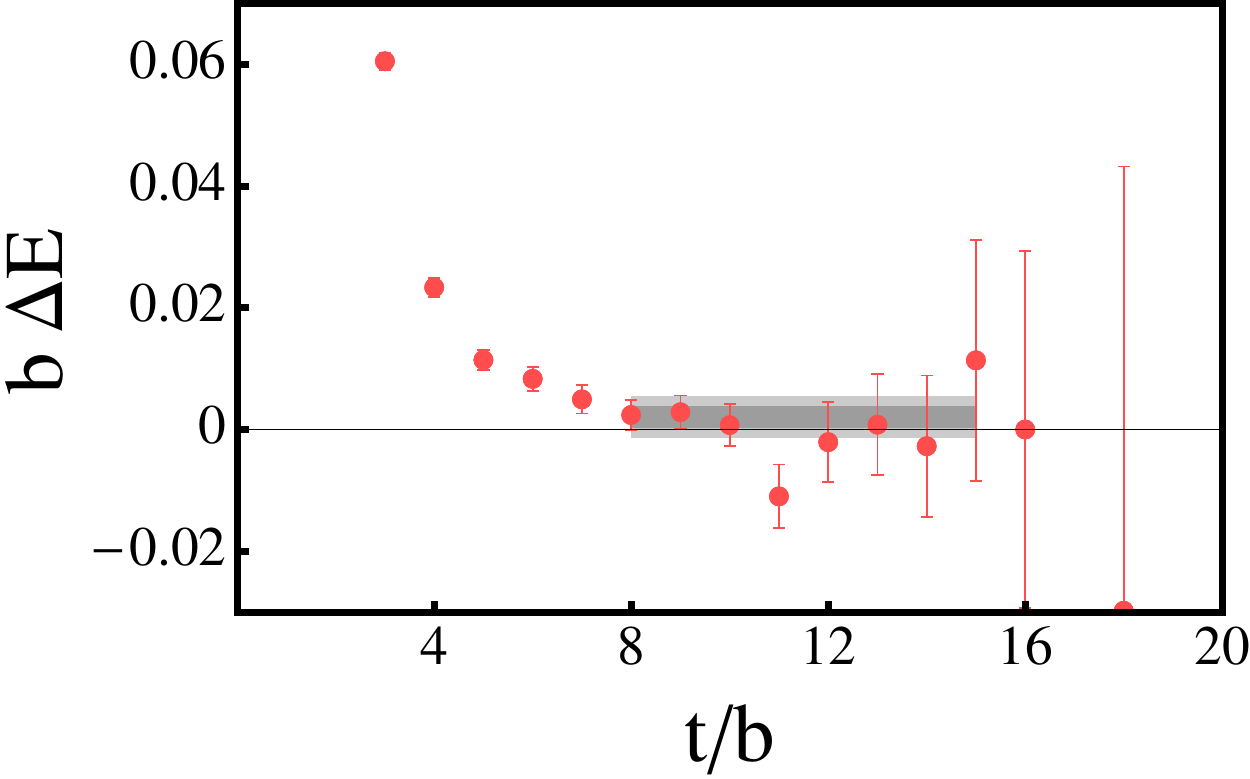}
  \caption{
The EMPs of the first excited states 
with  $|{\rm P}|=0$
in the $\si$ channel
in the $L=\Lafm$, $L=\Lbfm$ and $L=\Lcfm$ ensembles, respectively.
Twice the nucleon mass has been subtracted from the energy.
The dark (light) shaded regions correspond to the statistical uncertainty 
(statistical and systematic uncertainties combined in quadrature) of the fit to
the plateau over the indicated time interval.
 }
  \label{fig:1s0pzzero}
\end{figure}

\begin{figure}[!ht]
  \centering
  \includegraphics[width=0.32\textwidth]{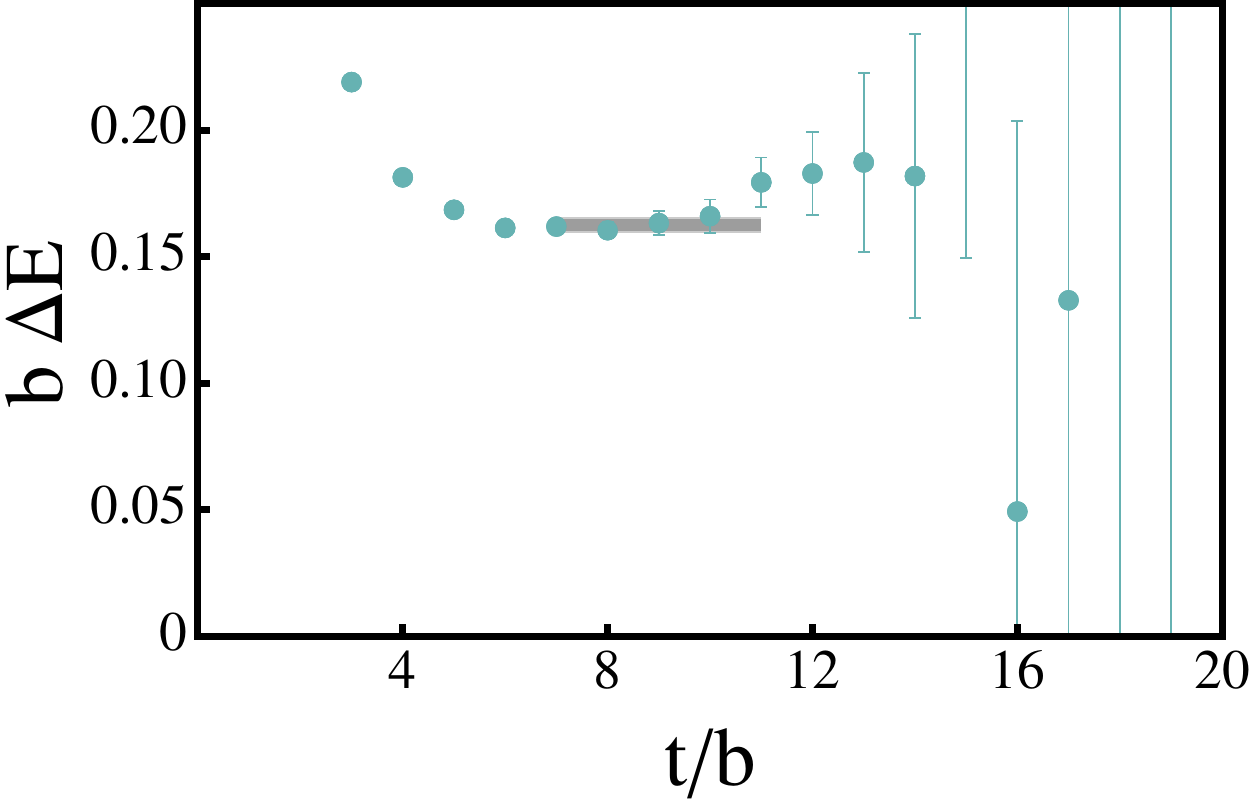}\,
  \includegraphics[width=0.32\textwidth]{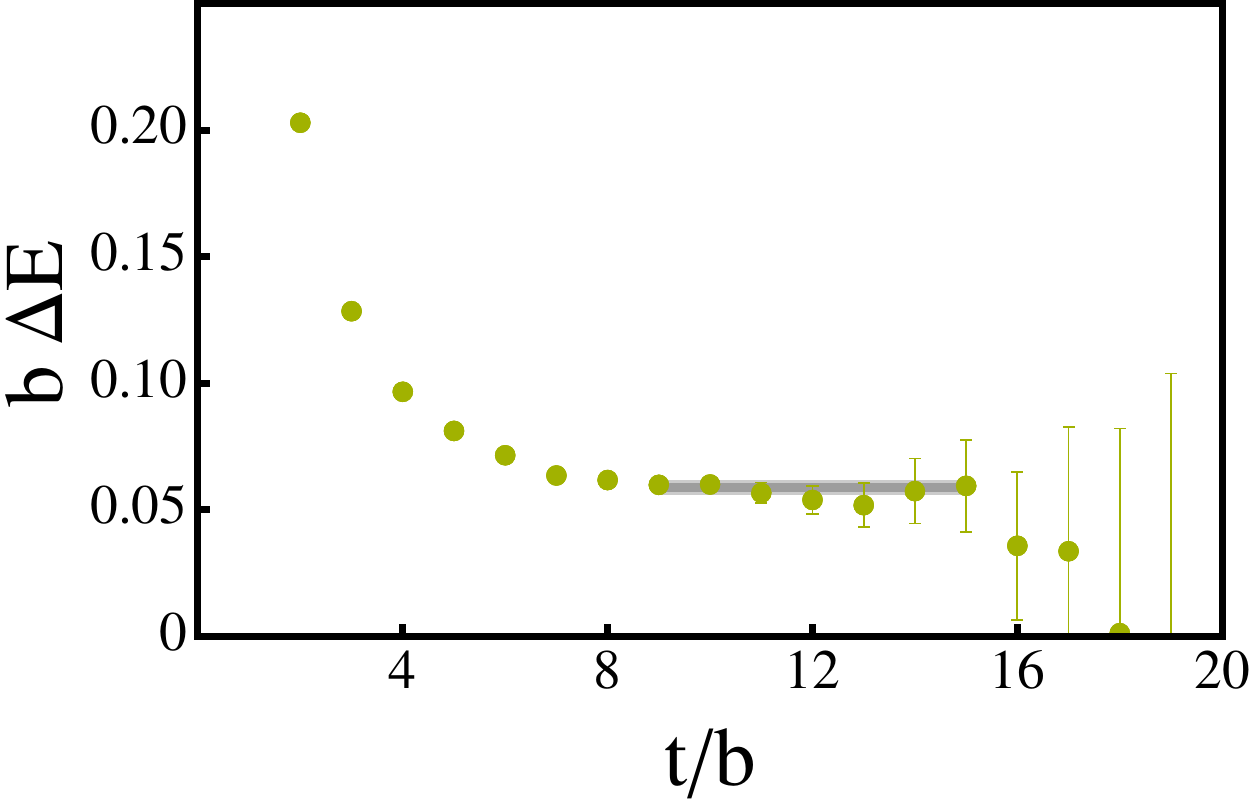}\,
  \includegraphics[width=0.32\textwidth]{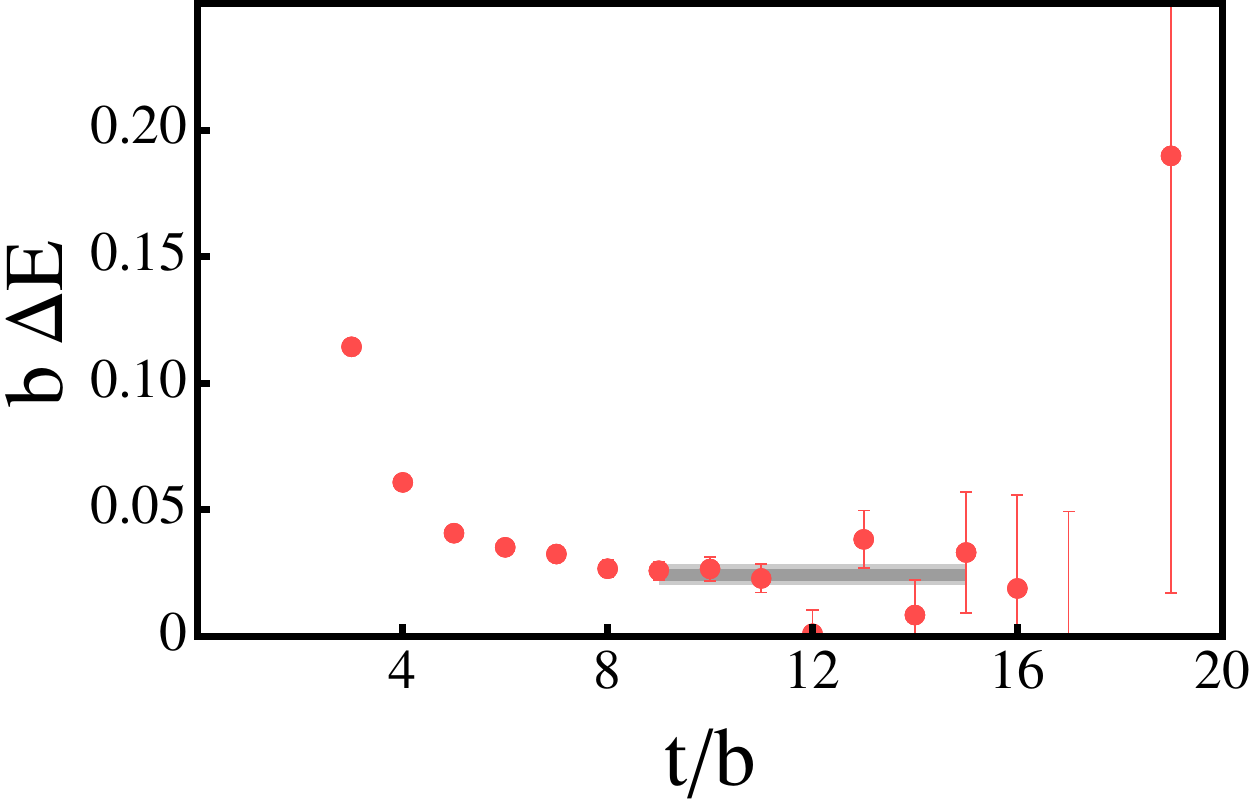} 
  \caption{
The EMPs of the first  excited state 
with  $|{\rm P}|=1$
in the $\si$ channel
in the $L=\Lafm$, $L=\Lbfm$ and $L=\Lcfm$ ensembles, respectively.
Twice the nucleon mass has been subtracted from the energy.
The dark (light) shaded regions correspond to the statistical uncertainty 
(statistical and systematic uncertainties combined in quadrature) of the fit to
the plateau over the indicated time interval.
 }
  \label{fig:1s0pzone}
\end{figure}

While the location of the bound state does not correspond to a real
value for the scattering phase shift, it does provide a real value of
$k\cot\delta$, and a valuable constraint on the scattering parameters
in the effective range expansion, which is valid below the t-channel
cut starting at $|{\bf k}|=m_\pi/2$.  Positively shifted energy
eigenvalues of two-body systems in the lattice volumes can be related
to scattering phase shifts using L\"uscher's
relation~\cite{Luscher:1986pf,Luscher:1990ux,Beane:2003da} and its extension to
boosted
systems~\cite{Rummukainen:1995vs,Kim:2005gf,Christ:2005gi,Davoudi:2011md}
provided they lie below the inelastic threshold.  The effective mass plots
(EMPs) associated with the two-nucleon energy, reduced by twice the
nucleon mass, of the first excited states with $|{\rm P}|=0$ are shown
in fig.~\ref{fig:1s0pzzero}.
Plateau's in the EMPs are found in all volumes, leading to clean
extractions of the energy splitting, and also to values of the phase
shift via L\"uscher's eigenvalue equation at those energies in the
$L=\Lafm$ and $L=\Lbfm$ ensembles.  The energy splitting in the
$L=\Lcfm$ ensemble is consistent with zero, and therefore straddles
the lowest singularity in the eigenvalue equation, thereby providing
no meaningful constraint on the phase shift or $k\cot\delta$.  The
energy splittings are determined with a correlated
$\chi^2$-minimization fit of a constant to the plateau in the EMPs
over the fit ranges shown in fig.~\ref{fig:1s0pzzero}.  Jackknife
resampling is used to generate the covariance matrix required to define
$\chi^2$ from the correlation functions.  Consistent results for the
extracted values of the energies are obtained using both one- and
two-state exponential fits to the correlation functions.  In the
figures, the inner shaded region corresponds to the statistical
uncertainty in the fit, derived from the $\chi^2$-minimization, and
the outer uncertainty corresponds to the fitting systematic
uncertainty combined in quadrature with the statistical uncertainty.
The fitting systematic uncertainty is determined by varying the
fitting interval over the extent of the plateau region and
accommodating the range of central fit values.  The EMPs associated
with the first excited state with $|{\rm P}|=1$ are shown in
fig.~\ref{fig:1s0pzone}.
The correlation functions in the $L=\Lbfm$ and $\Lcfm$ ensembles are
not sufficiently precise to provide a statistically meaningful phase
shift, as the energies are near singularities in L\"uscher's
eigenvalue equation.  This demonstrates a significant difficulty in determining
phase shifts in large volumes where the poles of the eigenvalue
equation are collapsing to form the infinite-volume scattering
continuum.  Table~\ref{tab:SingletResults} shows the results extracted
from the lowest-lying continuum levels from each of the ensembles.
\renewcommand{\arraystretch}{1.5}
\begin{table}
\begin{center}
\begin{minipage}[!ht]{16.5 cm}
  \caption{
Results from the lowest-lying continuum states in the $\si$ channel.
  }  
\label{tab:SingletResults}
\end{minipage}
\setlength{\tabcolsep}{0.3em}
\begin{tabular}{c|ccccc}
\hline
      Ensemble & $|{\rm P}|$ & $b \Delta E$ & 
      $|{\bf k}|/m_\pi$ & $k\cot\delta/m_\pi$ & $\delta$ $(^o)$  \\
\hline
$\latta$ & 0 & 
0.0358(13)(16) &  0.3506(64)(78)   &
$0.175^{+.034}_{-0.031} {}^{+0.043}_{-0.036}$   & 63.4(3.8)(4.7)  \\

$\latta$ & 1 & 0.1609(16)(37)  &  0.7197(41)(93) &
$-0.30^{+0.07}_{-0.07} {}^{+0.15}_{-0.17}$           \
& -67(5)(11) \\

$\lattb$ & 0 & 
0.0165(13)(22)  
& 0.2373(92)(96)  & 
$0.030^{+0.031}_{-0.028} {}^{+0.057}_{-0.046}$  & 83(7)(13) 
\\
\hline
\end{tabular}
\begin{minipage}[t]{16.5 cm}
\vskip 0.0cm
\noindent
\end{minipage}
\end{center}
\end{table}     

Below the inelastic threshold, at $|{\bf k}|^2= M_N m_\pi +
m_\pi^2/4$, where $k$ is the magnitude of the three-momentum of each
nucleon in the center-of-mass (CoM) frame, the s-wave scattering
amplitude can be uniquely described by a single phase shift and more
directly $k\cot\delta$.  Near threshold, and more generally, below the
t-channel cut, $k\cot\delta$ has a power-series expansion in terms of
the kinetic energy of the two-nucleons,
\begin{eqnarray}
k \cot\delta & = & -{1\over a}\ +\ {1\over 2} r |{\bf k}|^2\ +\  P |{\bf k}|^4\ +\ 
{\cal
  O}\left(|{\bf k}|^6 \right)
\ \ \ ,
\label{eq:ere}
\end{eqnarray}
called the effective range expansion (ERE), where $a$ is the
scattering length (using the nuclear physics sign convention), $r$ is
the effective range and $P$ is the shape parameter.  While the range
of possible values of the scattering length is unbounded, 
the size of the 
effective range and shape parameter are set by the range of the interaction.
\begin{figure}[!ht]
\begin{center}
\begin{minipage}[t]{4 cm}
\centerline{
\includegraphics[width=3.2in,angle=0]{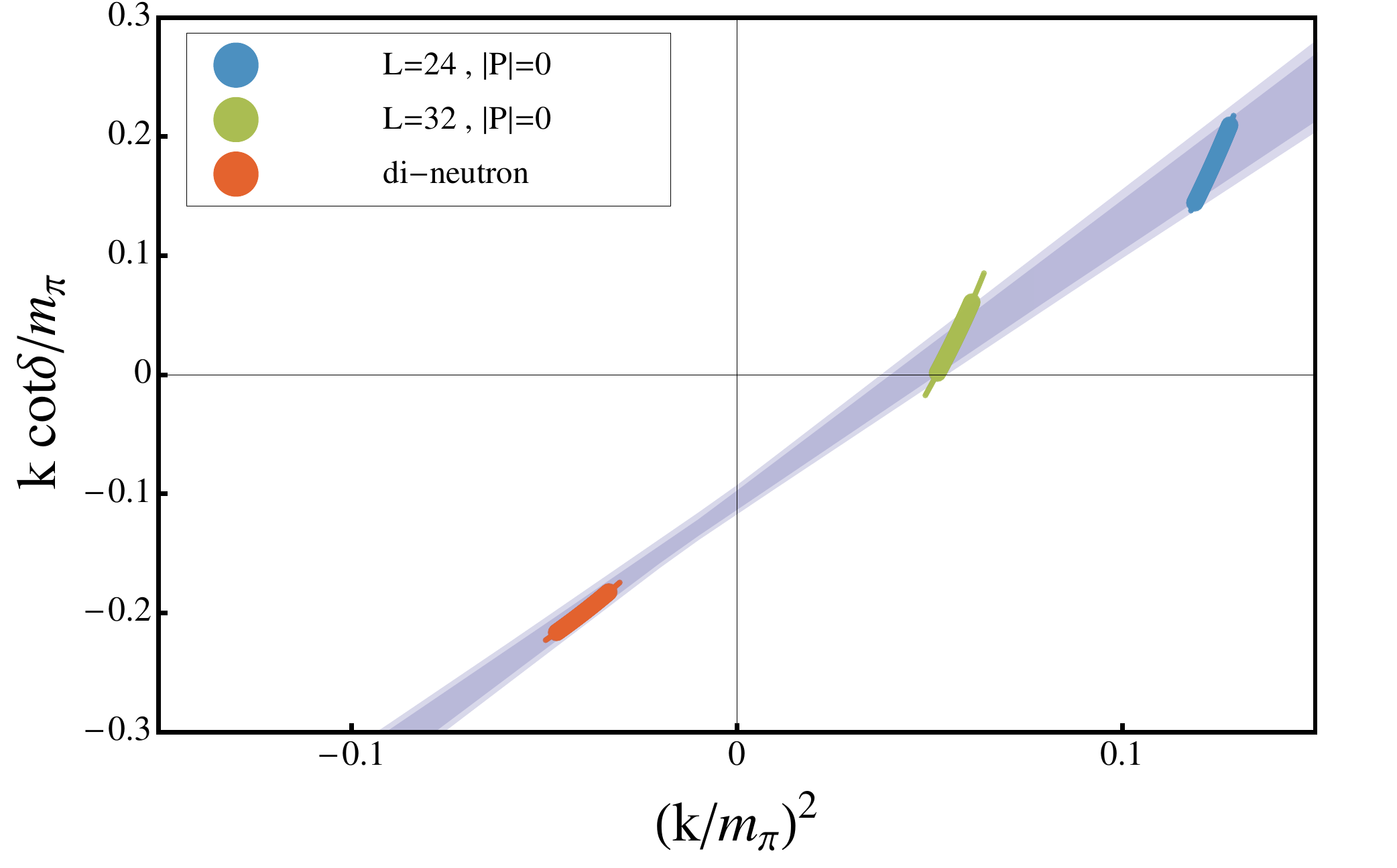}\quad
\includegraphics[width=3.2in,angle=0]{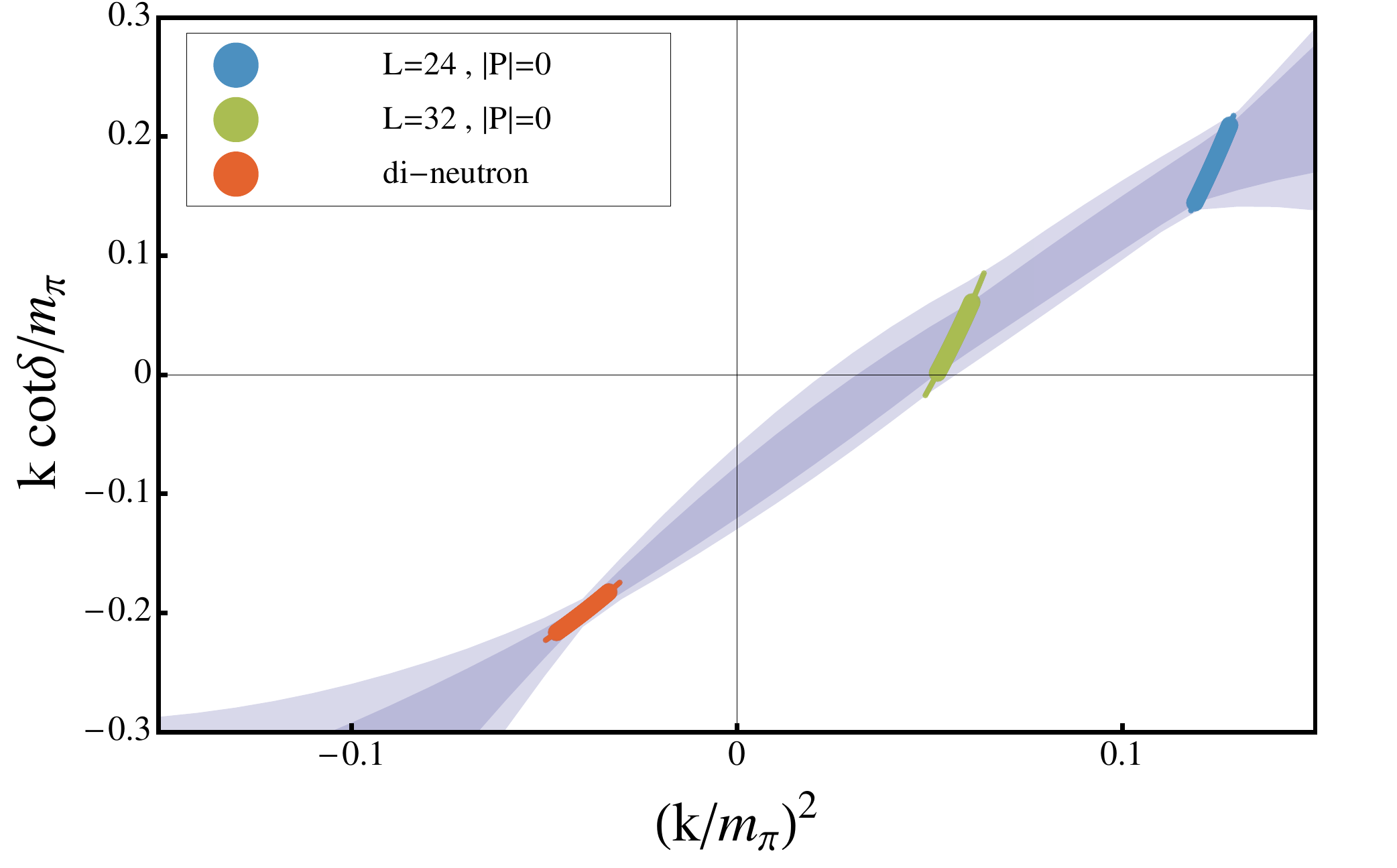}
}
\end{minipage}
\begin{minipage}[t]{16.5 cm}
\vspace{-0.1cm}
\caption{ 
$k\cot\delta$ in the $\si$ channel.
The positive energy values are given in Table~\ref{tab:SingletResults} and the
negative energy value is determined from the di-neutron binding energy.
The left panel is a two-parameter fit to the ERE, and the
right panel is a three-parameter fit to the ERE, as described in the text.
The inner (outer) shaded region corresponds to the statistical uncertainty 
(statistical and systematic uncertainties combined in quadrature).
}
 \label{fig:nnkcot}
\end{minipage}
\end{center}
\end{figure}
In fig.~\ref{fig:nnkcot}, the extracted values of $k\cot\delta/m_\pi$
given in Table~\ref{tab:SingletResults} for $|{\rm P}|=0$ and from the di-neutron
binding energy are shown as a function of $|{\bf k}|^2/m_\pi^2$.  The
three points shown in fig.~\ref{fig:nnkcot} lie significantly below
the t-channel cut and so the ERE of $k\cot\delta$ can be fit to define
the phase shift throughout this kinematic regime.  With three points
to fit, two-parameter (left panel) and three-parameter (right
panel) fits to the ERE of $k\cot\delta/m_\pi$ are performed and are
shown as the shaded regions in fig.~\ref{fig:nnkcot}.

\begin{figure}[!ht]
\begin{center}
\begin{minipage}[t]{4 cm}
\centerline{
\includegraphics[width=4.5in,angle=0]{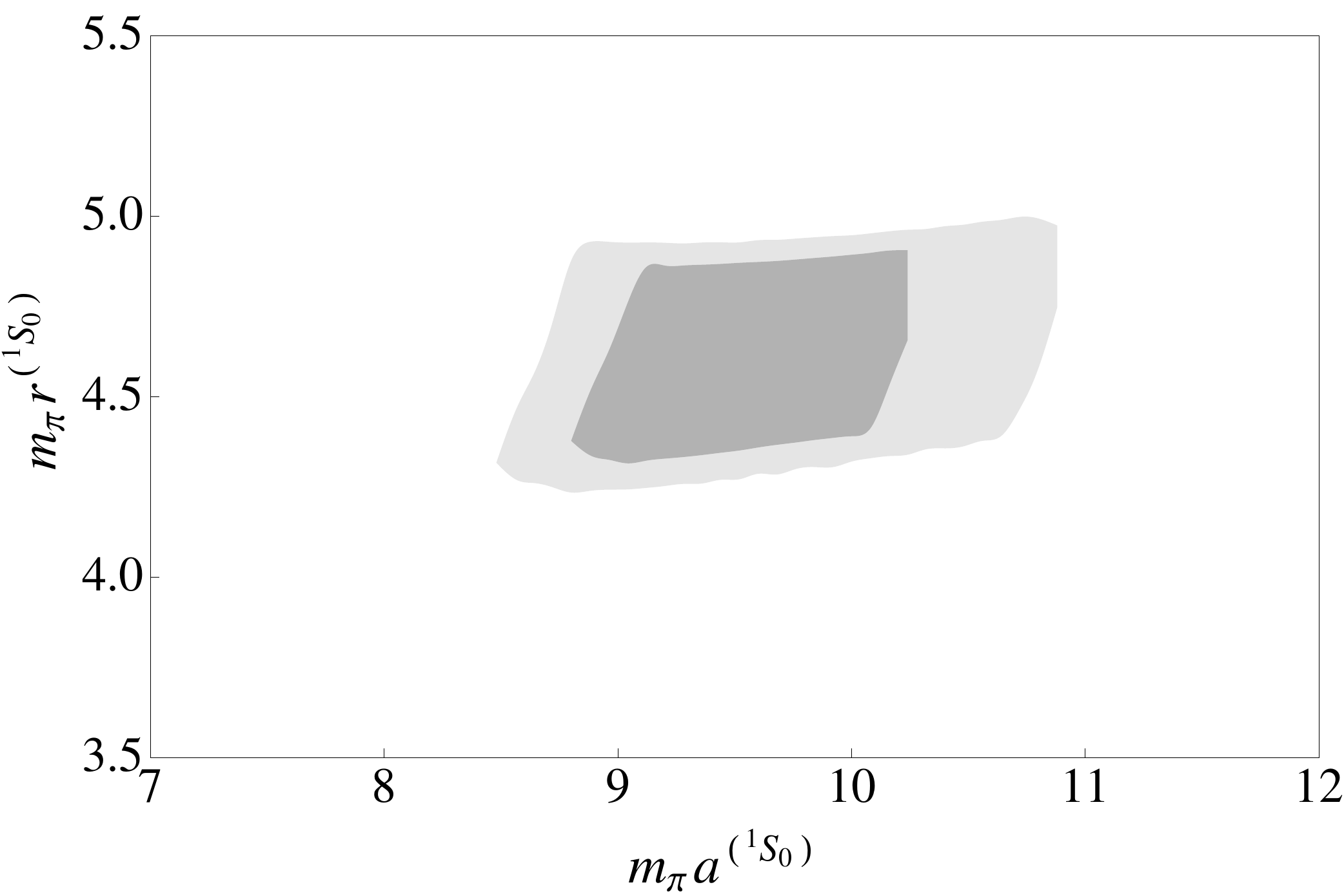}
}
\end{minipage}
\begin{minipage}[!ht]{16.5 cm}
\vspace{-0.1cm}
\caption{The 68$\%$ confidence region associated with 
$m_\pi a^{(\si)}$ and $m_\pi r^{(\si)}$ in the $\si$ channel.
The inner region corresponds to statistical uncertainties and the outer
region corresponds to statistical and systematic uncertainties combined in quadrature.
}
 \label{fig:nnamrm}
\end{minipage}
\end{center}
\end{figure}

The successful description by a two-parameter fit indicates small values of
the terms that are higher order in the ERE, consistent with what is
observed at the physical pion mass.  The scattering length and
effective range determined from the two-parameter fit are
\begin{eqnarray}
m_\pi a^{(\si)} & = & \ampising
\ \ \ ,\ \ \  
m_\pi r^{(\si)}  \ = \ \rmpising
\ \ \ ,
\end{eqnarray}
corresponding to
\begin{eqnarray}
a^{(\si)}  & = & \ampisingphys~{\rm fm}
\ \ \ , \ \ \ \
r^{(\si)}  \ = \ \rmpisingphys~{\rm fm}
\ \ \ .
\end{eqnarray}
The uncertainties associated with $a^{(\si)}$ and $r^{(\si)}$ are correlated,
and their 68$\%$ confidence region is shown in 
fig.~\ref{fig:nnamrm}.
The uncertainty in the scattering length is asymmetric as it is the
inverse scattering length that is the fit parameter. The shape parameter
obtained from the three parameter fit to the ERE expansion is consistent
with zero: $P m_\pi^3 = -1^{+4}_{-5}{}^{+5}_{-8}$. 
The scattering length
and effective range extracted from the three-parameter fit are consistent
with the two-parameter fit, but with larger uncertainties. A full quantification
of the theoretical error in the determination of the ERE parameters requires
more calculations than are currently available.

\begin{figure}[!ht]
\begin{center}
\begin{minipage}[t]{4 cm}
\centerline{
\includegraphics[width=3.2in,angle=0]{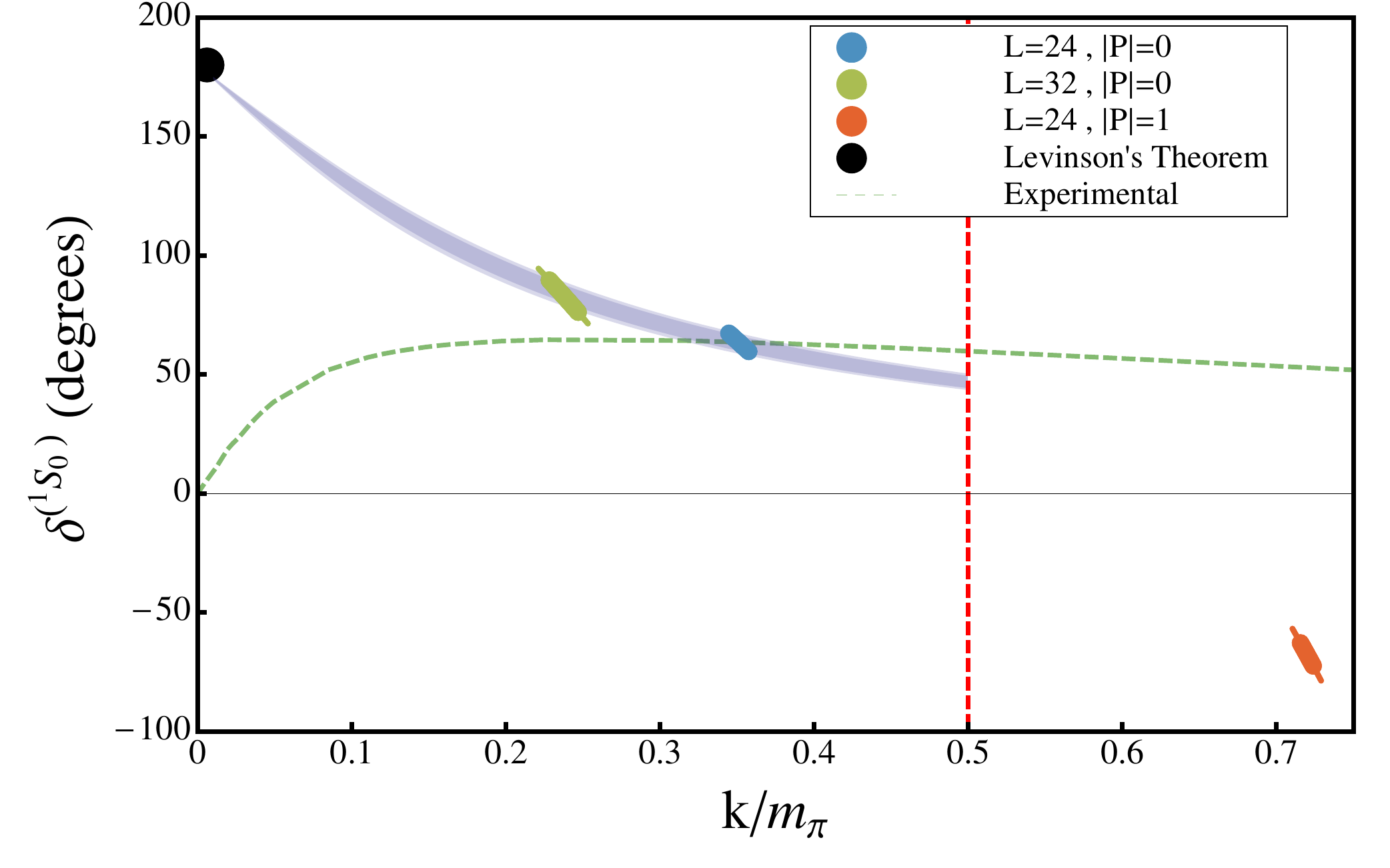}\quad
\includegraphics[width=3.2in,angle=0]{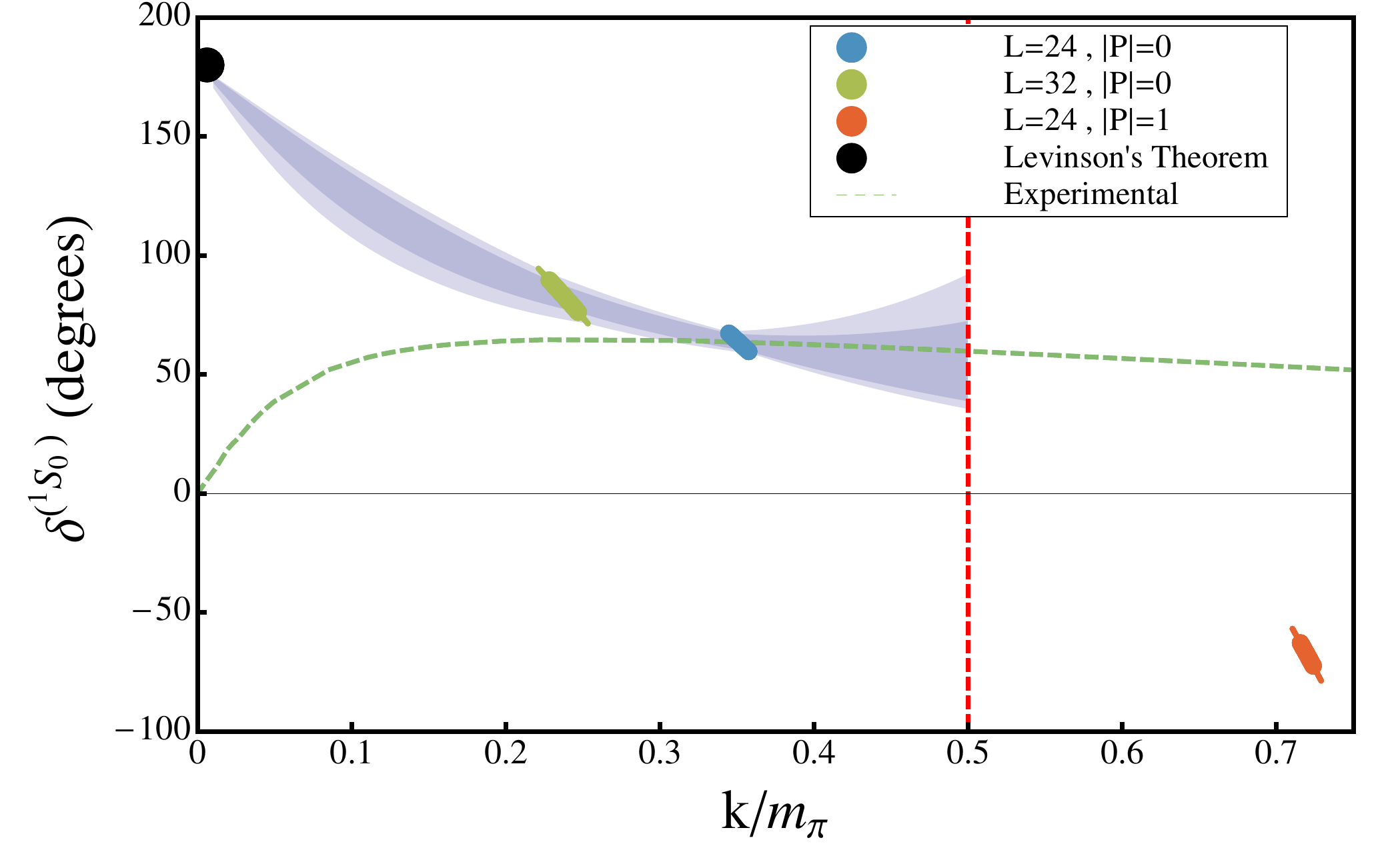}
}
\end{minipage}
\begin{minipage}[t]{16.5 cm}
\vspace{-0.1cm}
\caption{The phase shift in the $\si$ channel.
The left panel is a two-parameter fit to the ERE, 
while the right panel is a three-parameter fit to the ERE, as described in the text.
The inner (outer) shaded region corresponds to the statistical uncertainty 
(statistical and systematic uncertainties combined in quadrature)
in two- and three-parameter ERE fits to the results of the Lattice QCD calculation. 
The vertical (red) dashed line corresponds to the start of the t-channel cut
and the upper limit of the range of validity of the ERE.
The light (green) dashed line corresponds to the phase shift at the physical
pion mass  from the Nijmegen phase-shift analysis~\protect\cite{NNonline}.
}
 \label{fig:nndelta}
\end{minipage}
\end{center}
\end{figure}

The phase shift below the t-channel cut can be determined from these
fit parameters, and is shown in fig.~\ref{fig:nndelta}, along with the
results of the LQCD calculations and the phase shift at the physical
values of the quark masses.
We expect the phase shift predicted by the ERE to deviate
significantly from the true phase shift near the start of the
t-channel cut, and this is indeed suggested by
fig.~\ref{fig:nndelta}. Like the phase shift at the physical point,
the phase shift at the SU(3) symmetric point is found to change sign
at larger momenta, consistent with the presence of a repulsive hard
core in the NN interaction.

\section{Nucleon-Nucleon Scattering in the  $\siii-\diii$ Coupled Channels}
\label{sec:3s1}
\begin{figure}[!ht]
  \centering
  \includegraphics[width=0.32\textwidth]{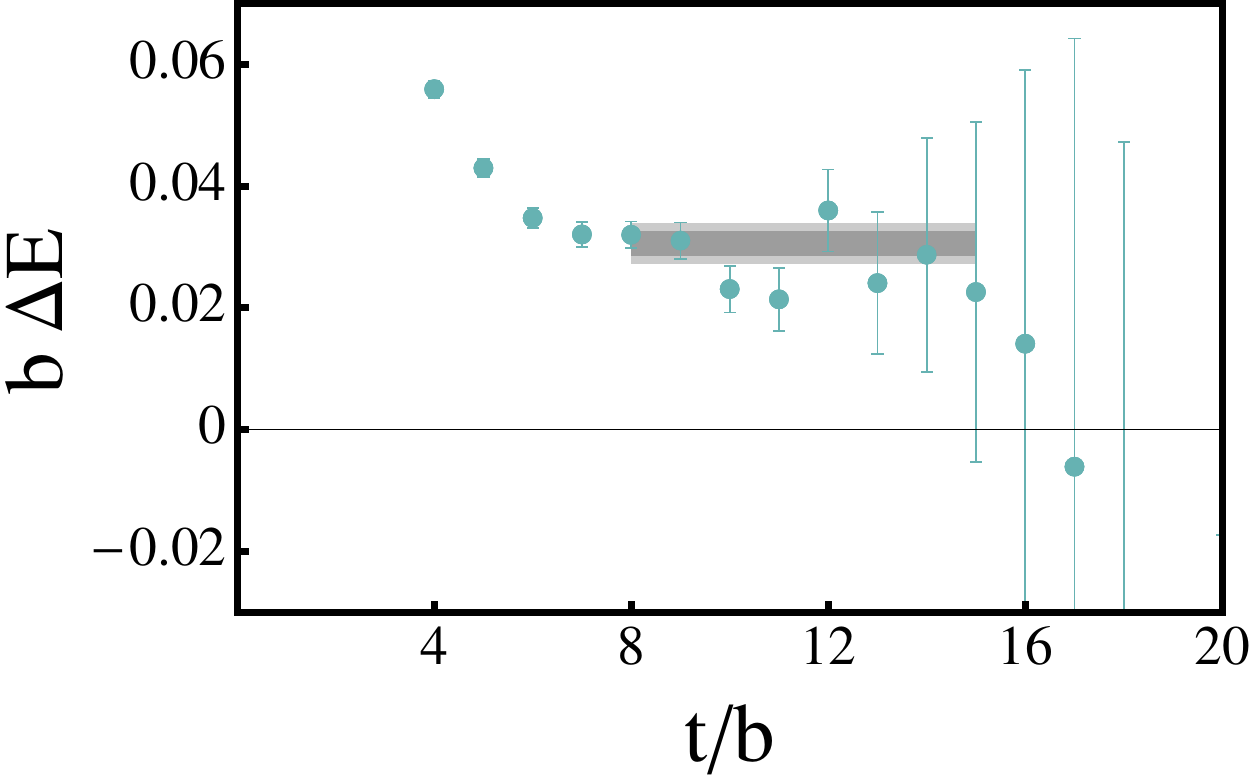}\,
  \includegraphics[width=0.32\textwidth]{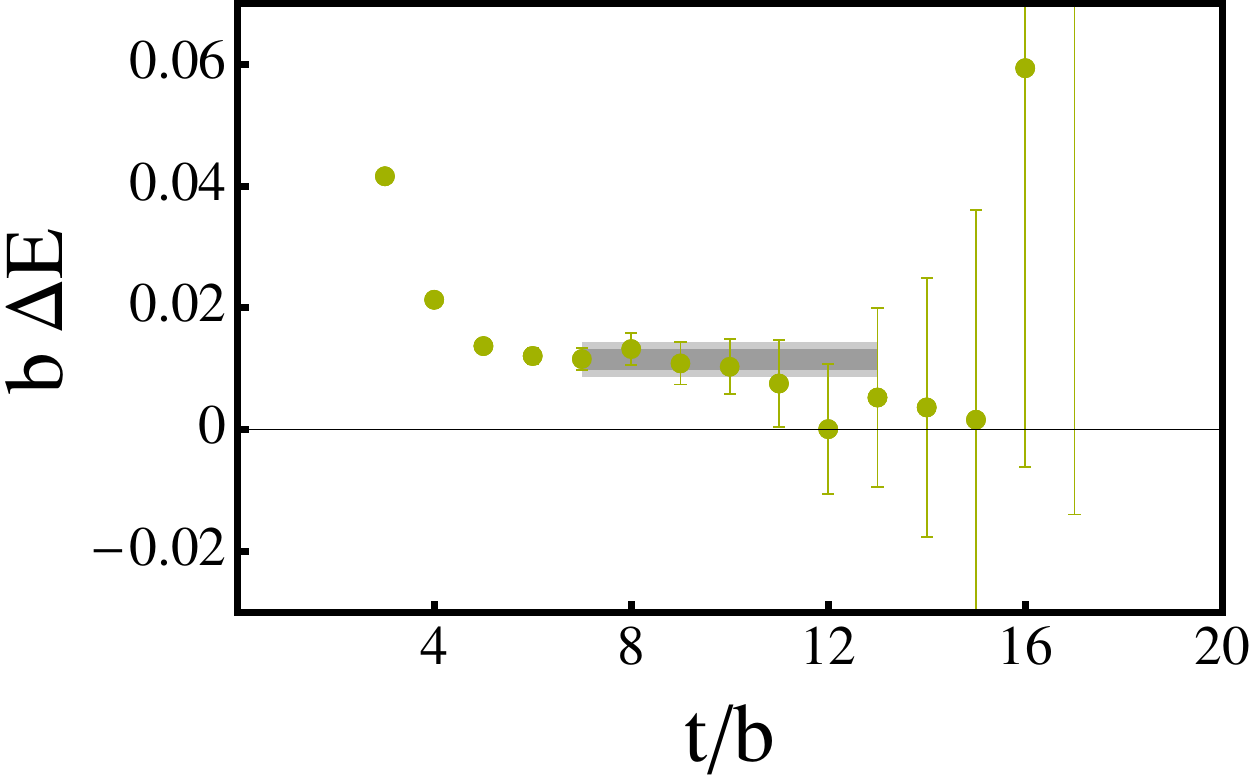}\,
  \includegraphics[width=0.32\textwidth]{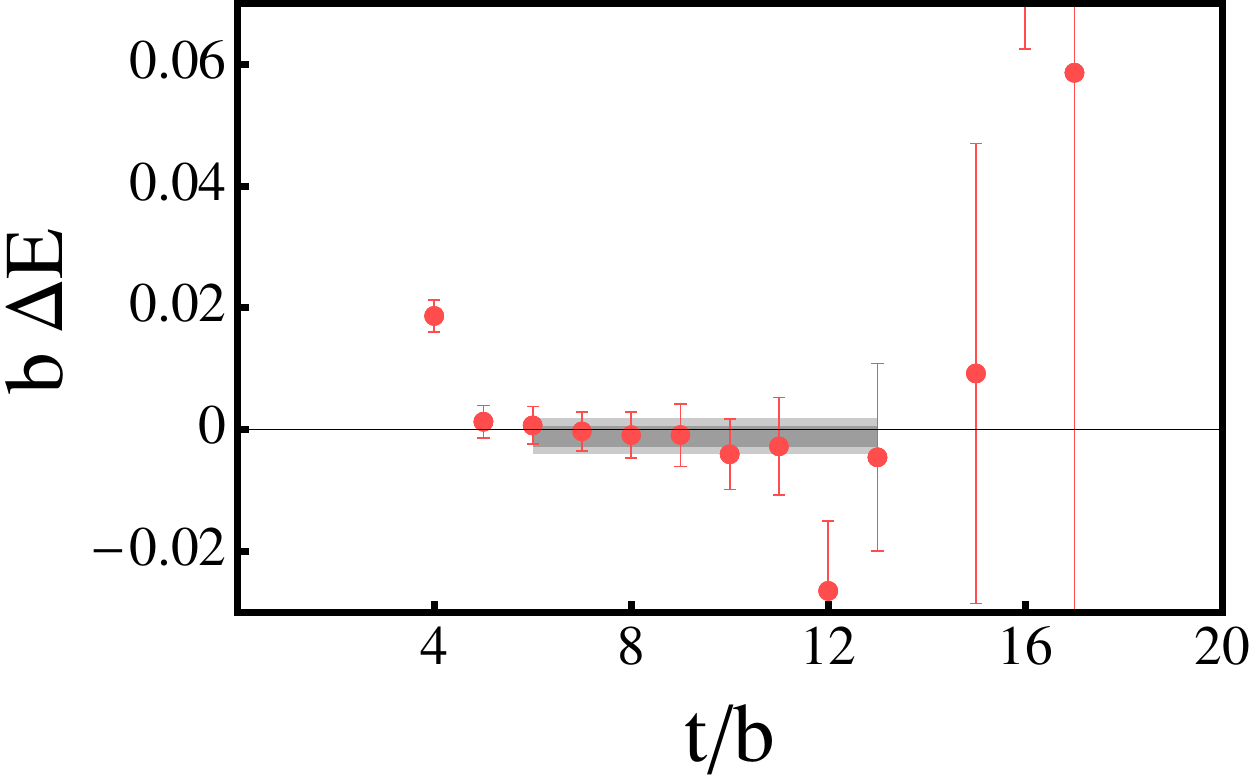}
  \caption{
The EMPs of the first excited states with $|{\rm P}|=0$ in the $\siii$ channel
in the $L=\Lafm$,  $L=\Lbfm$ and $L=\Lcfm$ ensembles, respectively.
Twice the nucleon mass has been subtracted from the energy.
The dark (light) shaded regions correspond to the statistical uncertainty 
(statistical and systematic uncertainties combined in quadrature) of the fit to
the plateau over the indicated time interval.
 }
  \label{fig:3s1pzzero}
\end{figure}

\begin{figure}[!ht]
  \centering
  \includegraphics[width=0.32\textwidth]{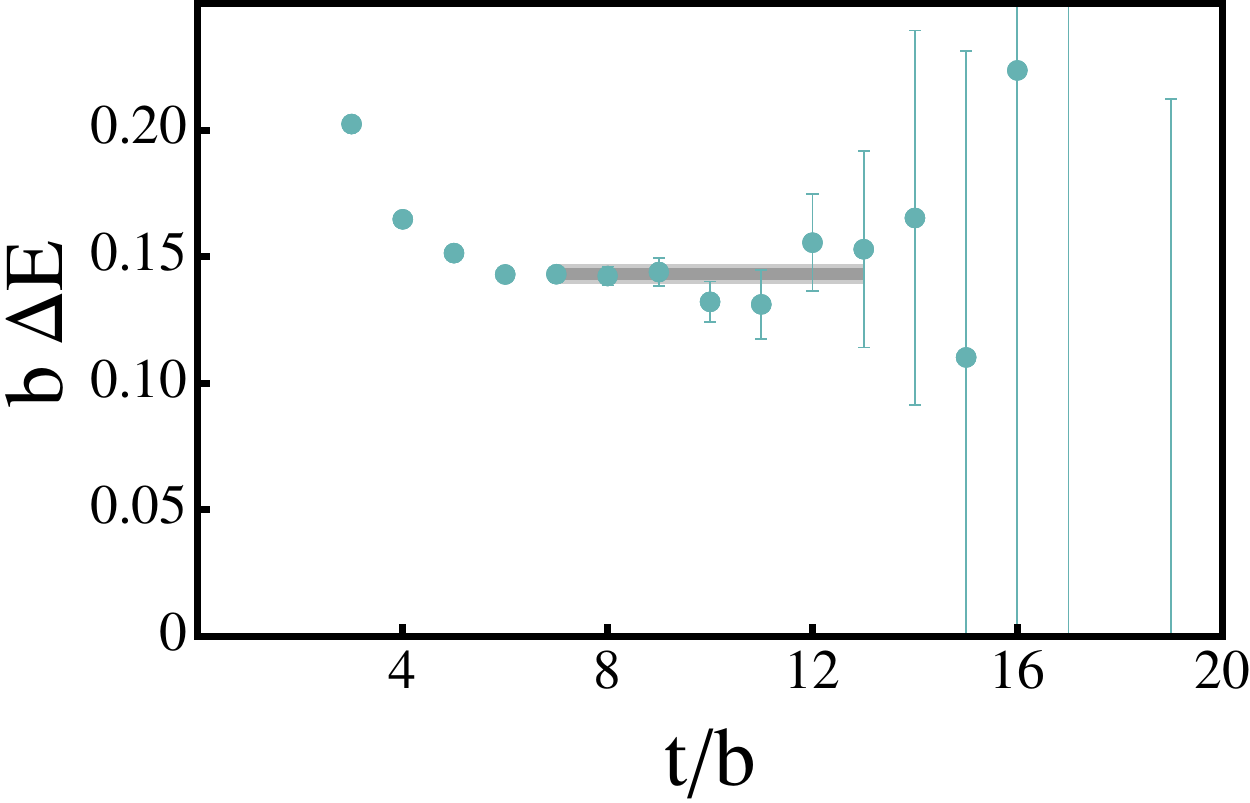}\,
  \includegraphics[width=0.32\textwidth]{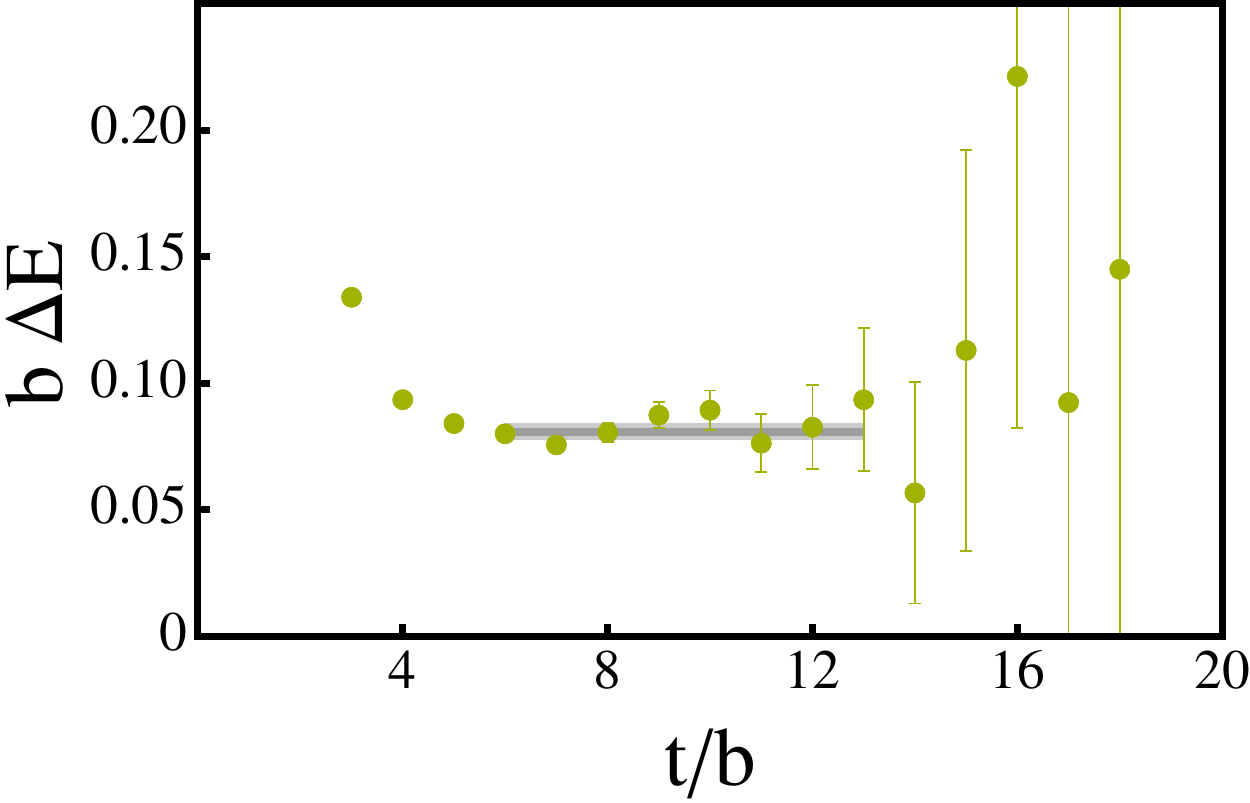}\,
  \includegraphics[width=0.32\textwidth]{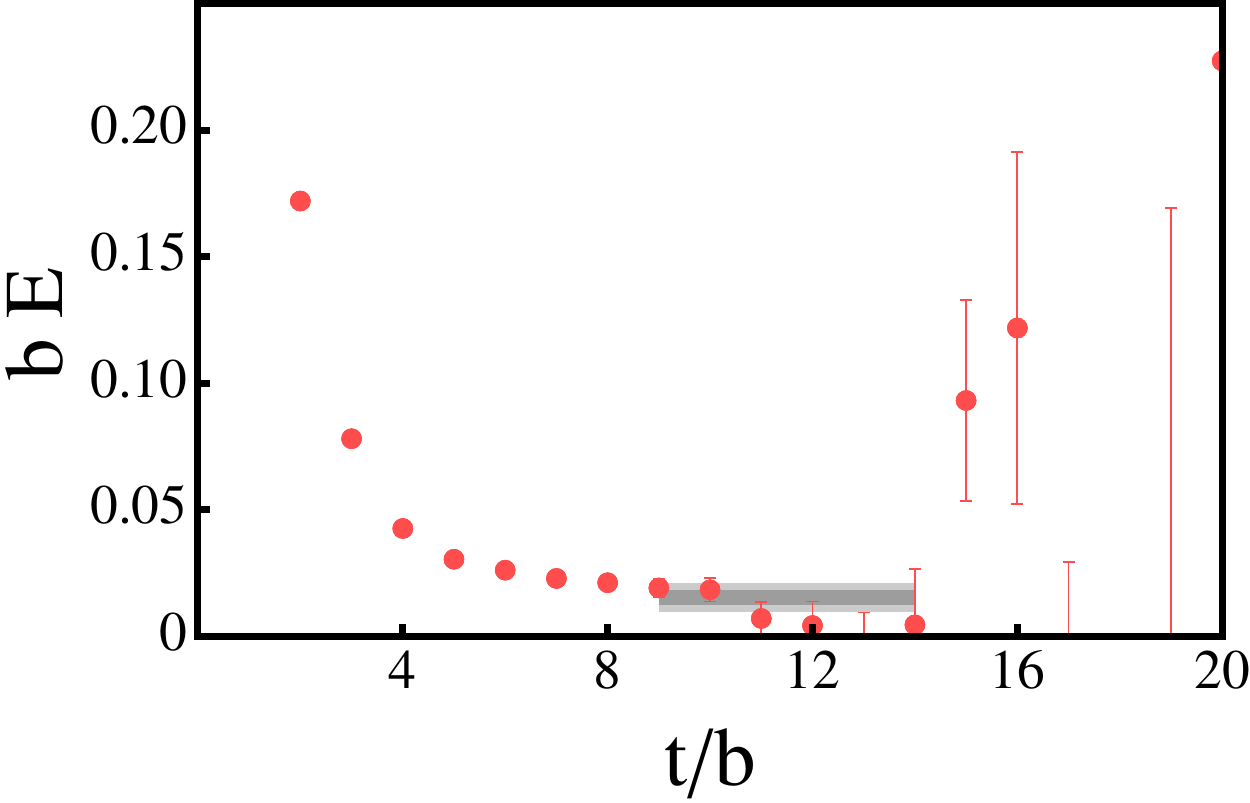} 
  \caption{
The EMPs of the first  excited states with $|{\rm P}|=1$
in the $\siii$ channel
in the $L=\Lafm$, $L=\Lbfm$ and $L=\Lcfm$ ensembles, respectively.
Twice the nucleon mass has been subtracted from the energy.
The dark (light) shaded regions correspond to the statistical uncertainty 
(statistical and systematic uncertainties combined in quadrature) of the fit to
the plateau over the indicated time interval.
 }
  \label{fig:3s1pzone}
\end{figure}

\begin{figure}[!ht]
\begin{center}
\begin{minipage}[t]{4 cm}
\centerline{
\includegraphics[width=4.5in,angle=0]{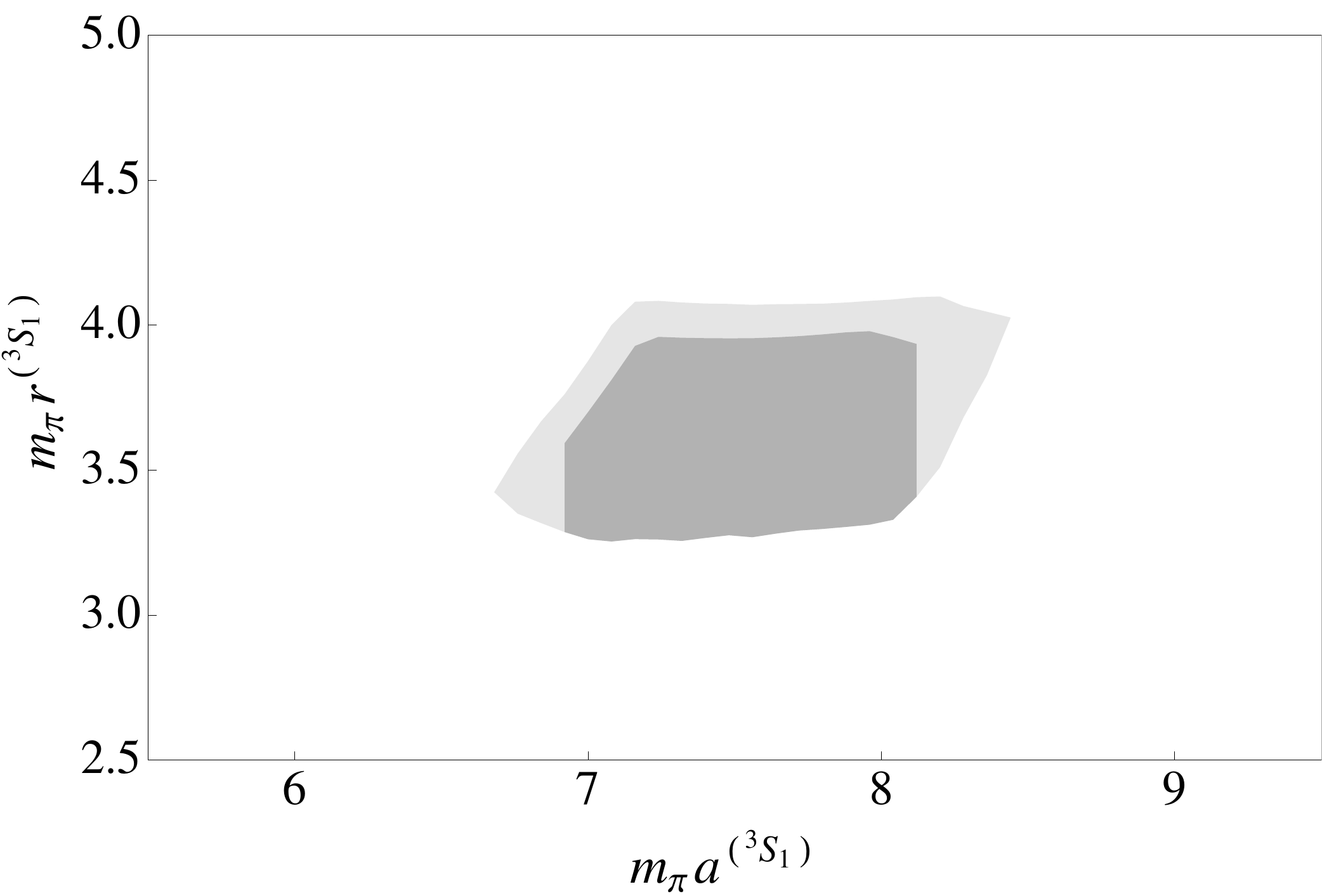}
}
\end{minipage}
\begin{minipage}[!ht]{16.5 cm}
\vspace{-0.1cm}
\caption{
The 68$\%$ confidence region associated with 
$m_\pi a^{(\siii)}$ and $m_\pi r^{(\siii)}$ in the $\siii$ channel.
The inner region corresponds to statistical uncertainties and the outer
region corresponds to statistical and systematic uncertainties combined in quadrature.
}
 \label{fig:deutamrm}
\end{minipage}
\end{center}
\end{figure}

\noindent
At the SU(3)-symmetric point, the deuteron is
bound~\cite{Beane:2012vq} by $B = \BdMeV$ which, as with the bound
di-neutron in the $\si$ channel, provides a constraint on
$k\cot\delta$. The deuteron has a d-wave component induced by the tensor
interaction, however mixing between the s-wave and the d-wave is
higher order in the ERE and first appears at the same order as the shape
parameter~\cite{Chen:1999tn}. Therefore, while the scattering length
and effective range are purely s-wave, the shape parameter is
contaminated by the d-wave admixture.  The EMPs associated with the
first excited states with $|{\rm P}|=0$ are shown in
fig.~\ref{fig:3s1pzzero}, and with $|{\rm P}|=1$ are shown in
fig.~\ref{fig:3s1pzone}.
The correlation functions calculated on the $L=\Lcfm$ ensemble have
energies that are too close to, or straddle, the singularities of L\"uscher's
eigenvalue equation and are not useful in determining the phase
shift. The results extracted from the fits to the plateau regions in
these EMPs are given in Table~\ref{tab:TripletResults}.
\renewcommand{\arraystretch}{1.5}
\begin{table}
\begin{center}
\begin{minipage}[!ht]{16.5 cm}
  \caption{
Results from the lowest-lying continuum states in the $\siii$ channel.
  }  
\label{tab:TripletResults}
\end{minipage}
\setlength{\tabcolsep}{0.3em}
\begin{tabular}{c|ccccc}
\hline
      Ensemble & $|{\rm P}|$ & $b \Delta E$ & 
      $|{\bf k}|/m_\pi$ & $k\cot\delta/m_\pi$ & $\delta$ $(^o)$  \\
\hline
$\latta$ & 0 & 0.0306(16)(23)  
& 0.324(8)(12)  
& $0.065^{+0.031}_{-0.029} {}^{+0.47}_{-0.40}$  
& 78.6(4.7)(6.9) 
\\
$\latta$ & 1 & 0.142(23)(20)  
& 0.6708(60)(54)  
& $-3.03^{+1.1}_{-4.0} {}^{+0.73}_{-9.3}$  &
$-12.5^{+7.1}_{-6.9} {}^{+6.5}_{-6.1} $
\\
$\lattb$ & 0 & 0.0115(17)(23)  
& 0.198(15)(19)  
& $-0.069(32)(43)$  
& 109(9)(13) 
\\
$\lattb$ & 1 
& 0.0788(24)(40)  
& 0.496(9)(14)  
&  -   
& -1(13)(22) 
\\
\hline
\end{tabular}
\begin{minipage}[t]{16.5 cm}
\vskip 0.0cm
\noindent
\end{minipage}
\end{center}
\end{table}     
\begin{figure}[!ht]
\begin{center}
\begin{minipage}[t]{4 cm}
\centerline{
\includegraphics[width=3.2in,angle=0]{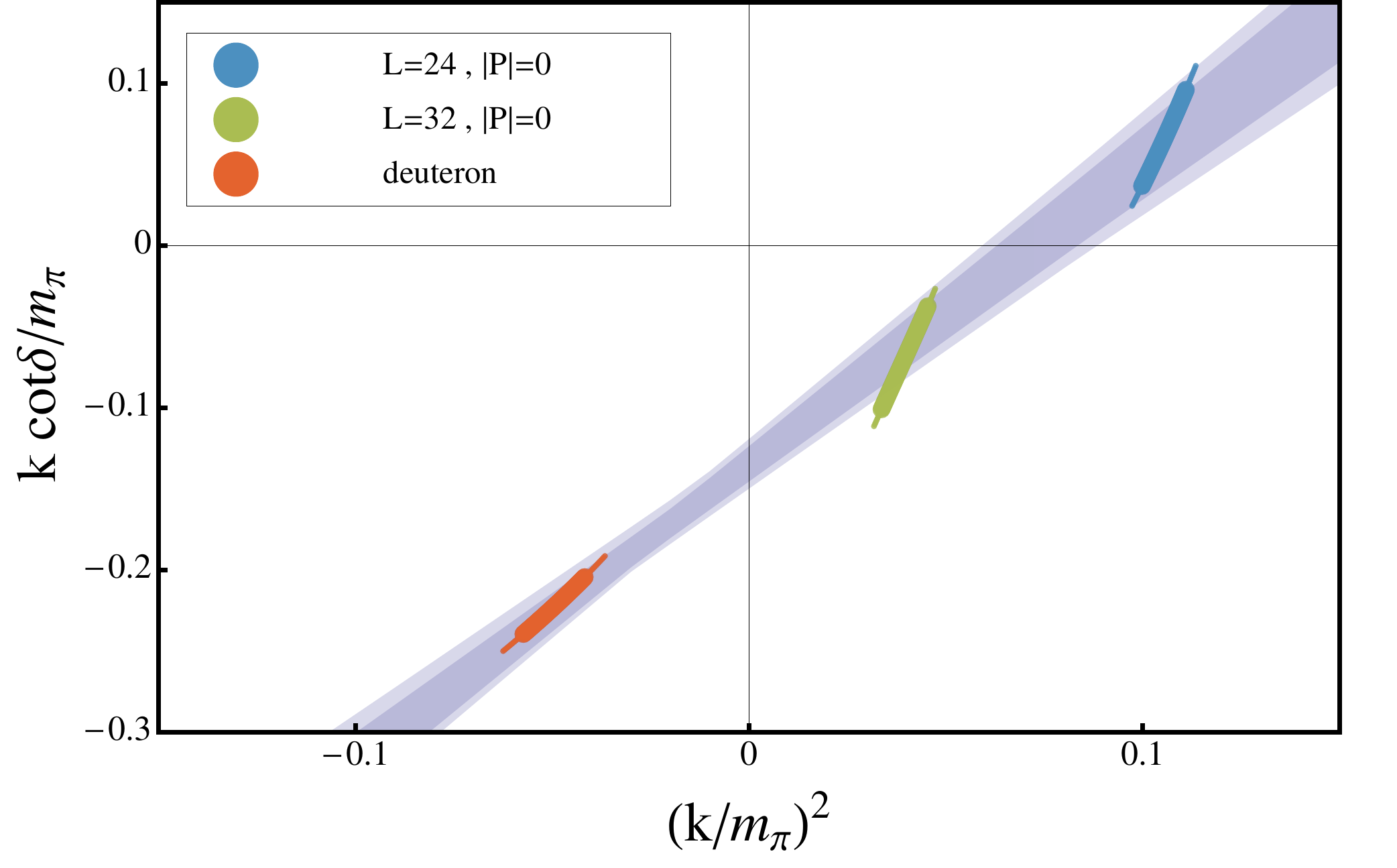}\quad
\includegraphics[width=3.2in,angle=0]{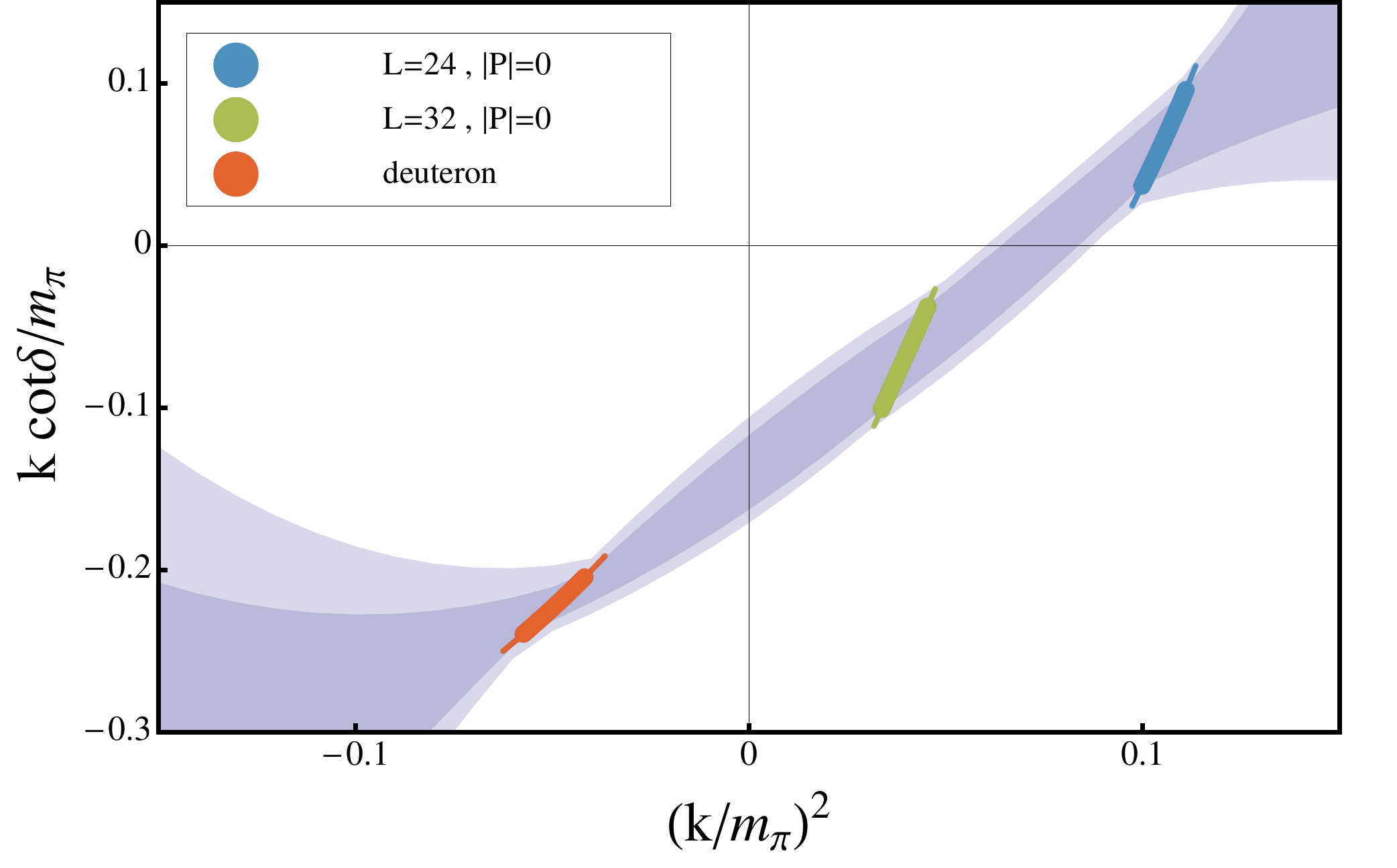}
}
\end{minipage}
\begin{minipage}[t]{16.5 cm}
\vspace{-0.1cm}
\caption{ 
$k\cot\delta$ in the $\siii$ channel.
The left panel is a two-parameter fit to the ERE, and the
right panel is a three-parameter fit to the ERE, as described in the text.
The positive energy values are given in Table~\ref{tab:TripletResults} and the
negative energy value is determined from the deuteron binding energy.
The inner (outer) shaded region corresponds to the statistical uncertainty 
(statistical and systematic uncertainties combined in quadrature).
}
 \label{fig:deutkcot}
\end{minipage}
\end{center}
\end{figure}
In fig.~\ref{fig:deutkcot}, the extracted values of
$k\cot\delta/m_\pi$ given in Table~\ref{tab:TripletResults} and from
the deuteron binding energy are shown as a function of $|{\bf
  k}|^2/m_\pi^2$. 
Following the procedure used to analyze the results in the $\si$-channel,
again with three points to fit, two-parameter (left panel) and
three-parameter (right panel) fits to the ERE of $k\cot\delta/m_\pi$
are performed and shown as the shaded regions in
fig.~\ref{fig:deutkcot}. The scattering
length and effective range determined from the two-parameter fit are
\begin{eqnarray}
m_\pi a^{(\siii)} & = & \ampitrip
\ \ \ ,\ \ \  
m_\pi r^{(\siii)}  \ = \ \rmpitrip
\ \ \ ,
\end{eqnarray}
corresponding to
\begin{eqnarray}
a^{(\siii)}  & = & \ampitripphys~{\rm fm}
\ \ \ , \ \ \ \
r^{(\siii)}  \ = \ \rmpitripphys~{\rm fm}
\ \ \ ,
\end{eqnarray}
and fig.~\ref{fig:deutamrm} 
shows the 68$\%$ confidence region for the extracted values of 
$a^{(\siii)}$ and $r^{(\siii)}$.
The shape parameter obtained from the three parameter fit to the ERE expansion is
consistent with zero: $P m_\pi^3 = 2^{+5}_{-6}{}^{+5}_{-6}$.
Again the scattering length and effective range extracted from the three-parameter fit are consistent
with the two-parameter fit, but with larger uncertainties. 

\begin{figure}[!ht]
\begin{center}
\begin{minipage}[t]{4 cm}
\centerline{
\includegraphics[width=3.2in,angle=0]{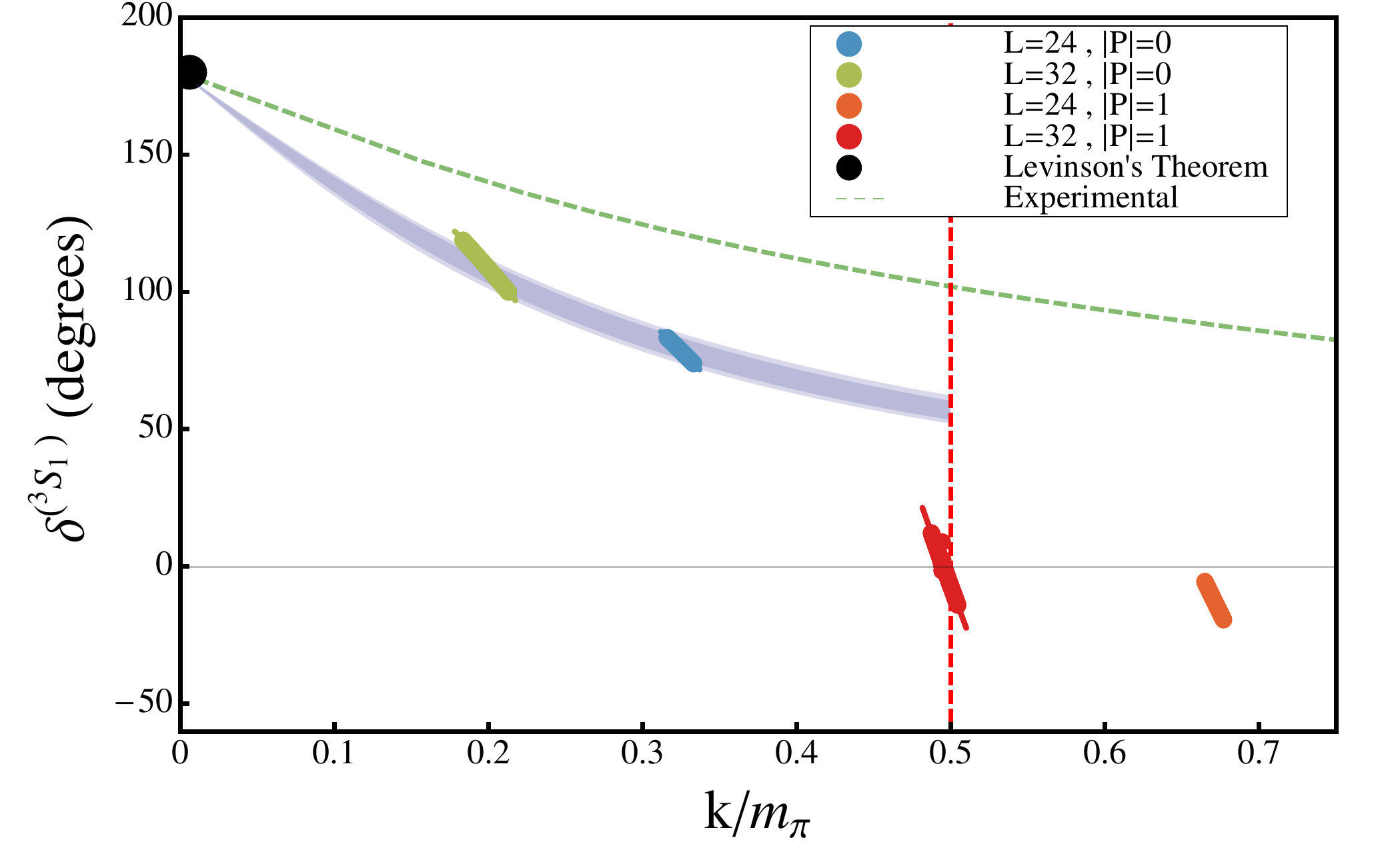}\quad
\includegraphics[width=3.2in,angle=0]{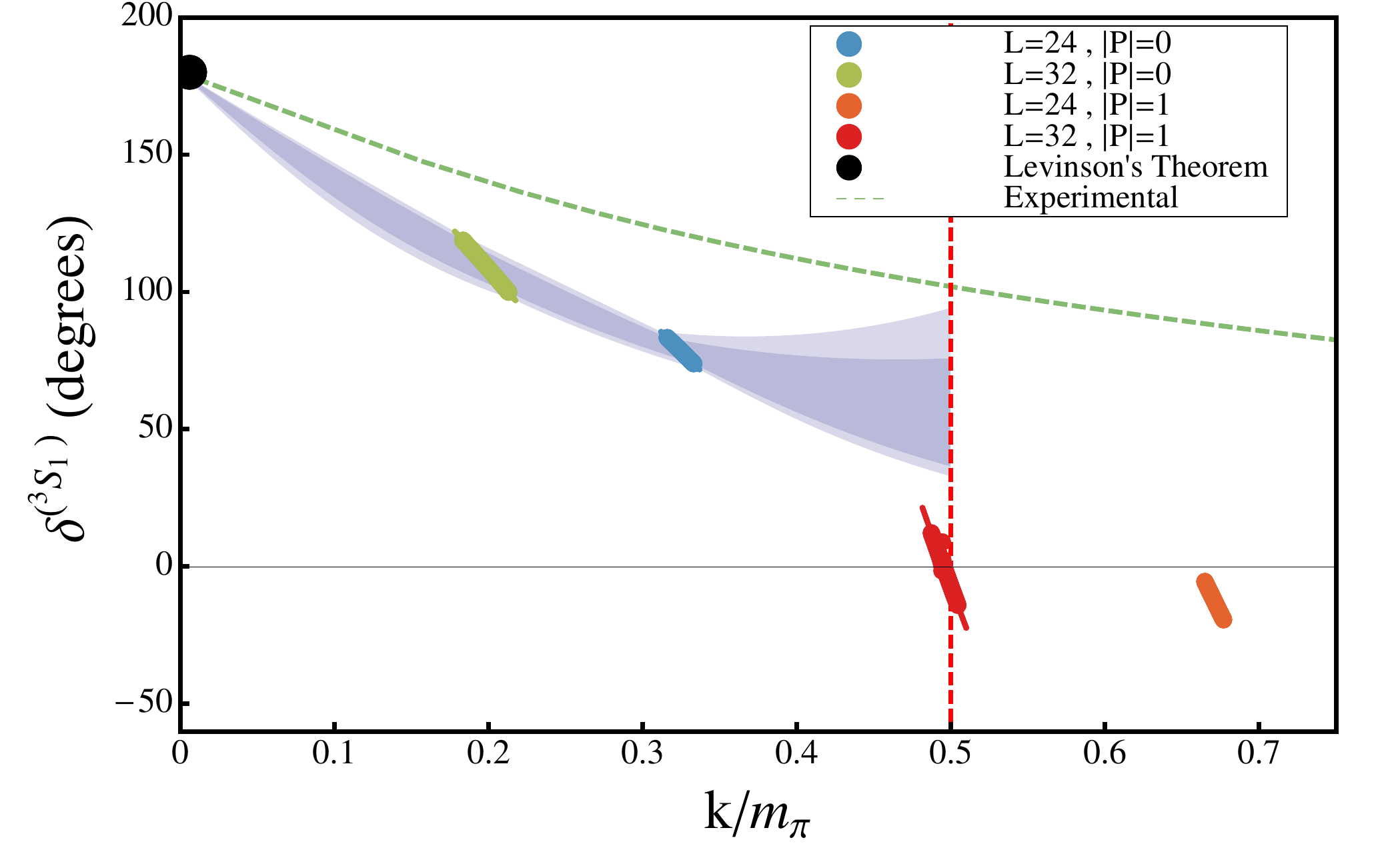}
}
\end{minipage}
\begin{minipage}[t]{16.5 cm}
\vspace{-0.1cm}
\caption{The phase shift in the $\siii$ channel.
The left panel is a two-parameter fit to the ERE, while the
right panel is a three-parameter fit to the ERE, as described in the text.
The inner (outer) shaded region corresponds to the statistical uncertainty 
(statistical and systematic uncertainties combined in quadrature)
in two- and three-parameter ERE fit to the results of the Lattice QCD calculation. 
The vertical (red) dashed line corresponds to the start of the t-channel cut
and the upper limit of the range of validity of the ERE.
The light (green) dashed line corresponds to the phase shift at the physical
pion mass from the Nijmegen phase-shift analysis~\protect\cite{NNonline}.
}
 \label{fig:deutdelta}
\end{minipage}
\end{center}
\end{figure}

The phase shift below the t-channel cut can be determined from these fit
parameters, and is shown in fig.~\ref{fig:deutdelta}, along with the results of
the LQCD calculations and the phase shift at the physical point.
As in the $\si$ channel, the phase shift predicted by the ERE is
expected to deviate significantly from the true phase shift near the
t-channel cut, and this is seen in fig.~\ref{fig:deutdelta}.
Like the $\siii$ phase shift at the physical point, and the phase
shift we have obtained in the $\si$ channel, the phase shift at the SU(3)
symmetric point is found to change sign at larger momenta, consistent
with the presence of a repulsive hard core in the NN
interaction.

\section{Nucleon-Nucleon Effective Ranges}
\label{sec:NNR}
\noindent
Unlike the scattering length, the size of the effective range and the
higher-order contributions to the ERE are set by the range of the
interaction.  The leading estimate of the effective range for light
quarks is $r\sim 1/m_\pi$, and higher order contributions are expected
to be suppressed by further powers of the light-quark masses.  It is
natural to consider an expansion of the product $m_\pi r$ in the
light-quark masses.  While the most general form of the expansion
contains terms that are non-analytic in the pion
mass~\cite{Weinberg:1990rz,Weinberg:1991um,Kaplan:1998tg,Kaplan:1998we}, for instance of the form $m_q
\log m_q$, with determinations at only two pion masses (including the
experimental value) a polynomial fit function is chosen,
\begin{eqnarray}
m_\pi r & = & 
A \ +\ B\ m_\pi\ +\ ...
\ \ \ .
\label{eq:rfitform}
\end{eqnarray}
\begin{figure}[!ht]
\begin{center}
\begin{minipage}[t]{4 cm}
\centerline{
\includegraphics[width=3.0in,angle=0]{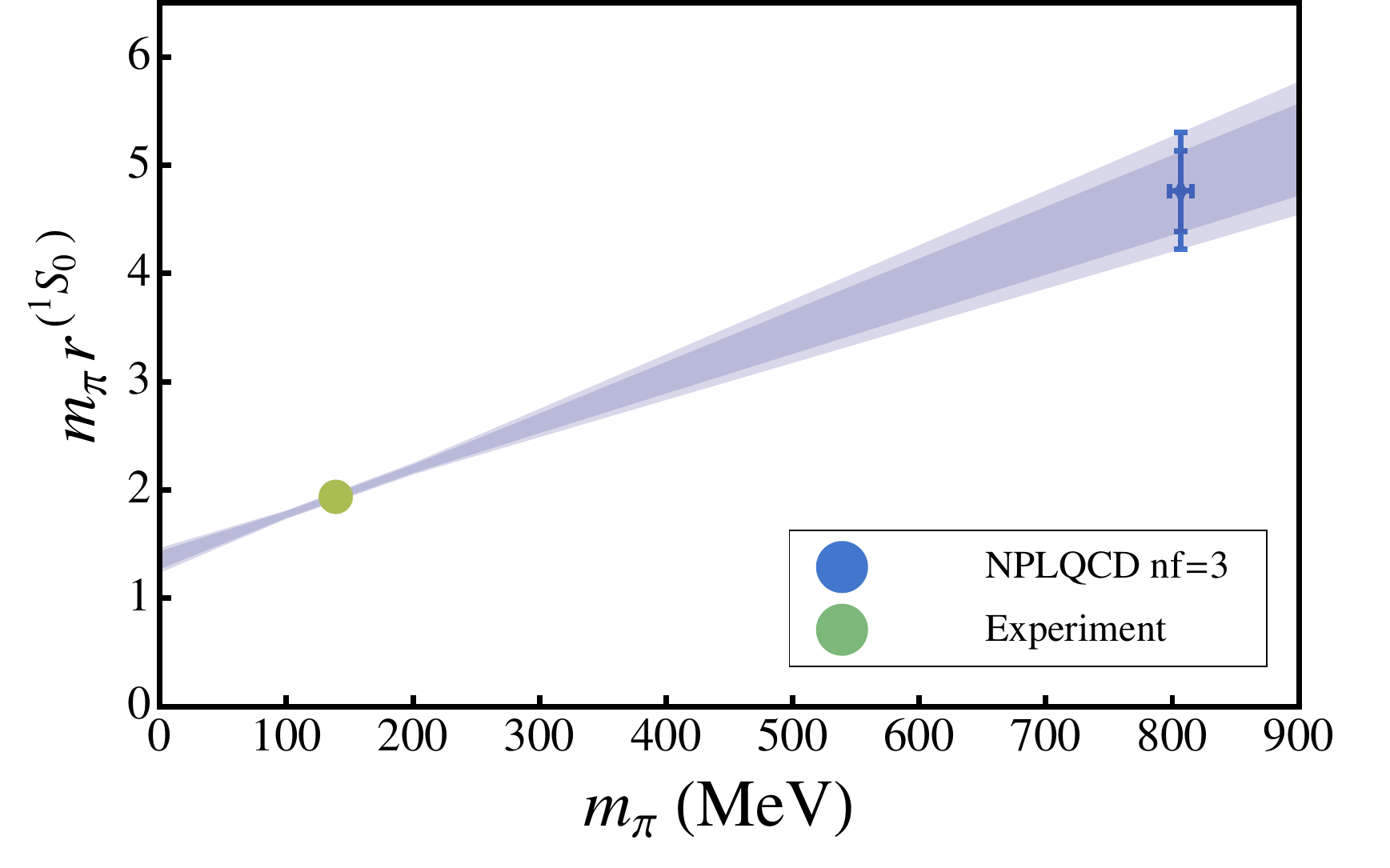}\ \ \ \ 
\includegraphics[width=3.0in,angle=0]{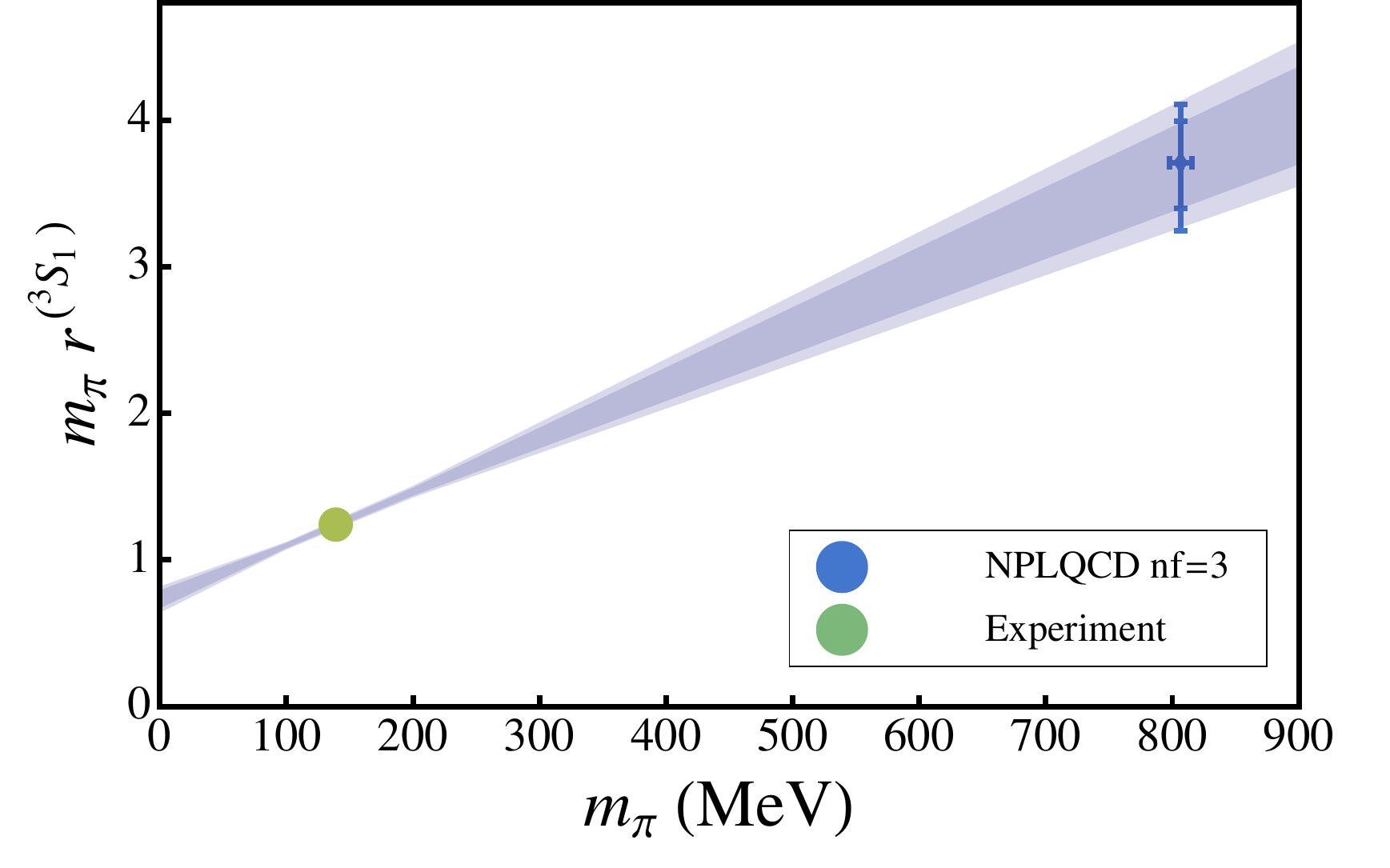}
}
\end{minipage}
\begin{minipage}[t]{16.5 cm}
\vspace{-0.1cm}
\caption{The NN effective range in the $\si$ channel (left panel)
  and the $\siii$ channel (right panel).
The inner (outer) shaded region corresponds to the statistical uncertainty 
(statistical and systematic uncertainties combined in quadrature)
in a two-parameter fit to the results of the Lattice QCD calculation and the experimental value. 
}
 \label{fig:NNmr}
\end{minipage}
\end{center}
\end{figure}
In fig.~\ref{fig:NNmr}, the results of our LQCD calculations of $m_\pi
r$ are shown, along with the experimental value in each channel and a
fit to the form given in eq.~(\ref{eq:rfitform}). While the
uncertainties in the lattice determinations are somewhat large
compared to those of the experimental determination, it appears that
there is modest dependence upon the light-quark masses.  The fit
values are
\begin{eqnarray}
A^{(\si)} & = & 1.348^{+0.080}_{-0.080} {}^{+0.079}_{-0.083}
\ \ ,\ \ 
B^{(\si)} \ =\ 4.23^{+0.55}_{-0.56} {}^{+0.59}_{-0.57}~{\rm GeV}^{-1}
\nonumber\\
A^{(\siii)} & = & 0.726^{+0.065}_{-0.059} {}^{+0.072}_{-0.059}
\ \ ,\ \ 
B^{(\siii)} \ =\ 3.70^{+0.42}_{-0.47} {}^{+0.42}_{-0.52}~{\rm GeV}^{-1}
\ \ \ .
\end{eqnarray}
The two-parameter fit is clearly over simplistic and more precise LQCD calculations are required at smaller light-quark
masses to better constrain the light-quark mass dependence of the effective
ranges.

\section{Fine Tunings and SU(4) Spin-Flavor Symmetry}
\label{sec:FTs}
\noindent
At the physical values of the quark masses, the deuteron is an
interesting system as it is much larger than the range of the nuclear
force.  Its binding energy is determined by the pole in the scattering
amplitude in the $\siii-\diii$ coupled channels.  It is known very
precisely at the physical light-quark masses, $B_d=2.224644(34)~{\rm
  MeV}$, and recently LQCD calculations of the deuteron binding have
been performed at unphysical light-quark
masses~\cite{Yamazaki:2011nd,Beane:2011iw,Yamazaki:2012hi,Beane:2012vq}.
Given that both the scattering lengths and effective ranges calculated
in this work are large compared with the pion Compton wavelength
(which naively dictates the range of the interaction for light pions),
we explore the naturalness of the two-nucleon systems. In this context,
naturalness is defined by the length scales of the system as compared
to the range of the interaction.  By contrast, a fine-tuned quantity
is one in which the length scales of the system are unnatural over a
small range of parameters of the underlying theory.

\begin{figure}[!ht]
\begin{center}
\begin{minipage}[t]{4 cm}
\centerline{
\includegraphics[width=3.2in,angle=0]{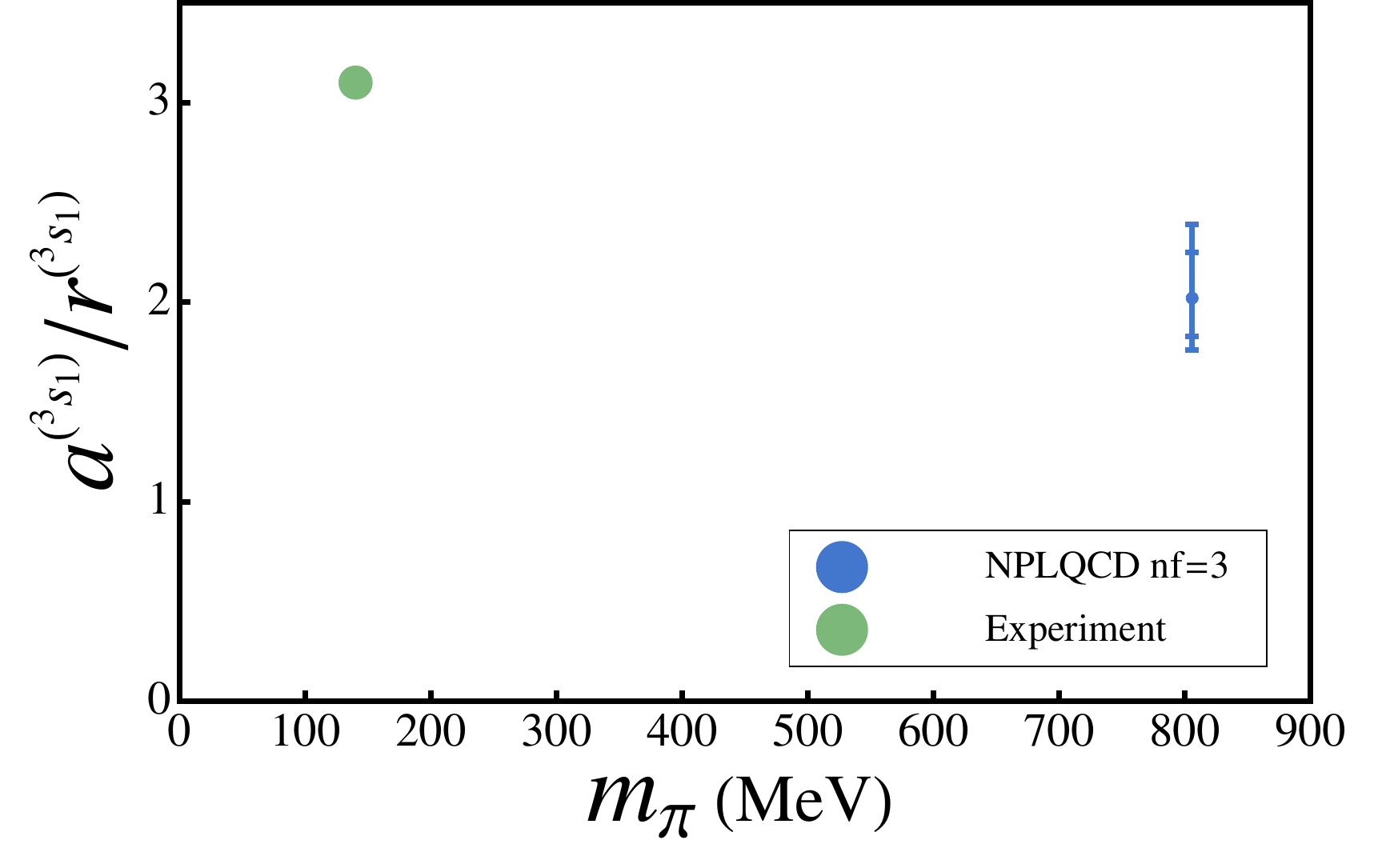}\ \ \
\includegraphics[width=3.2in,angle=0]{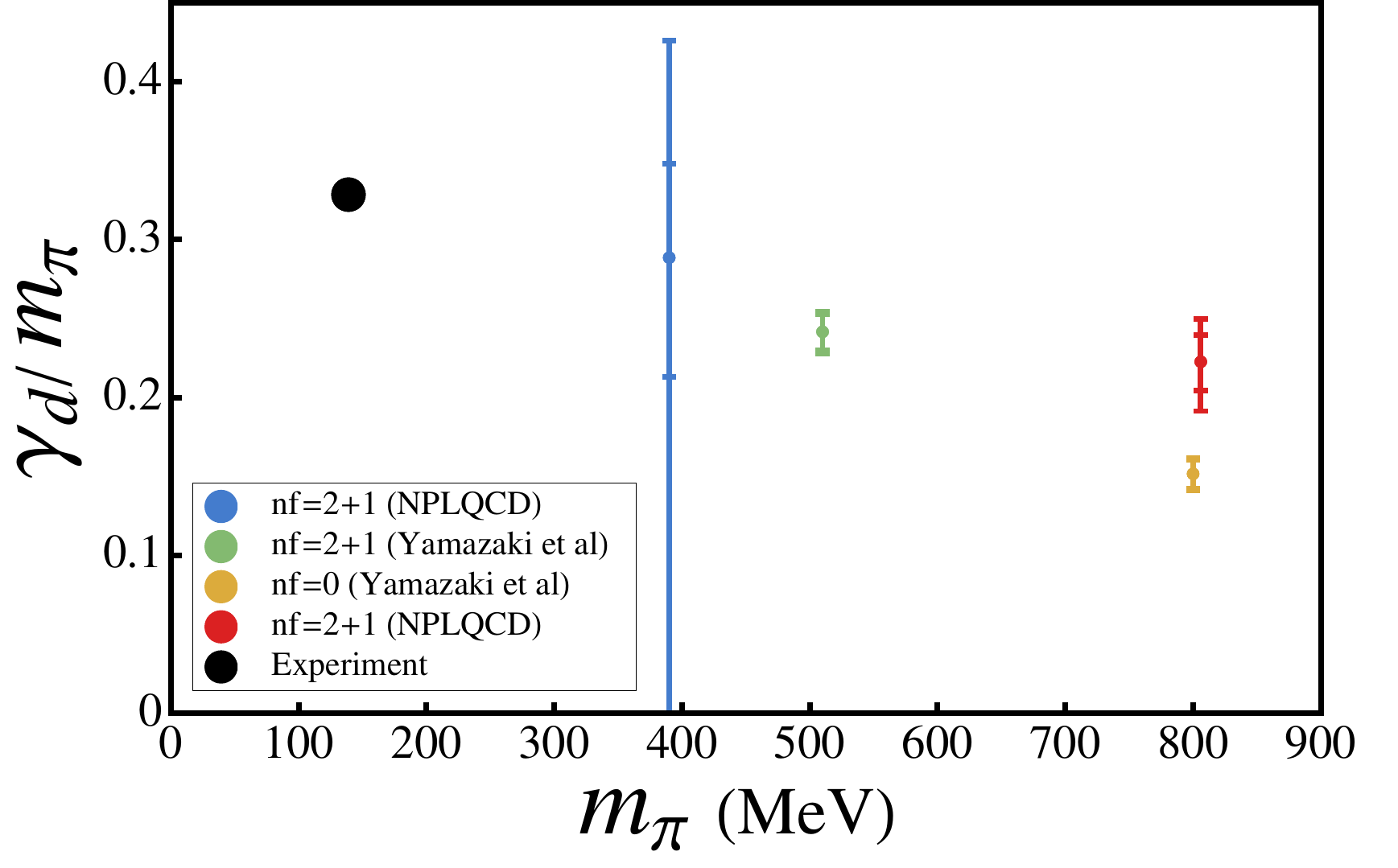}
}
\end{minipage}
\begin{minipage}[t]{16.5 cm}
\vspace{-0.1cm}
\caption{
The left panel shows the ratio of the scattering length to effective range in
the $\siii$ channel.
The right panel shows the normalized deuteron binding momentum 
versus the pion mass~\cite{Yamazaki:2011nd,Beane:2011iw,Beane:2012vq,Yamazaki:2012hi}.
The black point denotes the experimental value.
}
 \label{fig:deutgamma}
\end{minipage}
\end{center}
\end{figure}

The left panel of fig.~\ref{fig:deutgamma} gives the ratio of the
scattering length to effective range in the $\siii$ channel as a
function of the pion mass.  As the effective range is a measure of the
range of the interaction, this figure reveals that the deuteron is
becoming more natural at heavier light-quark masses.  In the right
panel of fig.~\ref{fig:deutgamma}, the deuteron binding momentum
$\gamma_d$ (related to the binding energy by $B_d = \gamma_d^2/M_N$)
normalized to the pion mass is shown as a function of the pion
mass. In the chiral regime one would expect that that $\gamma_d$
scales as $m_\pi^2$ as suggested by effective field
theory~\cite{Beane:2001bc,chiralextrapol_seattle,chiralextrapol_bonn,seattle,Braaten:2003eu,victorandbob,chen,CarrilloSerrano:2012ja}.
However, at the heavy up and down quark masses used here, naive
expectations based on the uncertainty principle suggest that the
deuteron binding momentum, if natural, would scale roughly as the
inverse of the range of the interaction.  As the ratio of $\gamma_d$
to $m_\pi$ as a function of $m_\pi$ is not constant, but rather is
falling, we conclude that pion exchange is no longer the only
significant contribution to the long-range component of the nuclear
force, consistent with the meson spectrum found at these quark masses.

While more precise calculations at these quark masses are desirable, and
LQCD calculations at other light-quark masses and at other lattice
spacings are required to make definitive statements, the present
calculations suggest that the deuteron remains unnatural over a large
range of light-quark masses. This would imply that the unnaturalness
of the deuteron binding energy at the physical point is a generic
feature of QCD with three light quarks and does not result from a
fine-tuning of their masses.  If subsequently confirmed, this would be
a very interesting result.

\begin{figure}[!ht]
\begin{center}
\begin{minipage}[t]{4 cm}
\centerline{
\includegraphics[width=3.2in,angle=0]{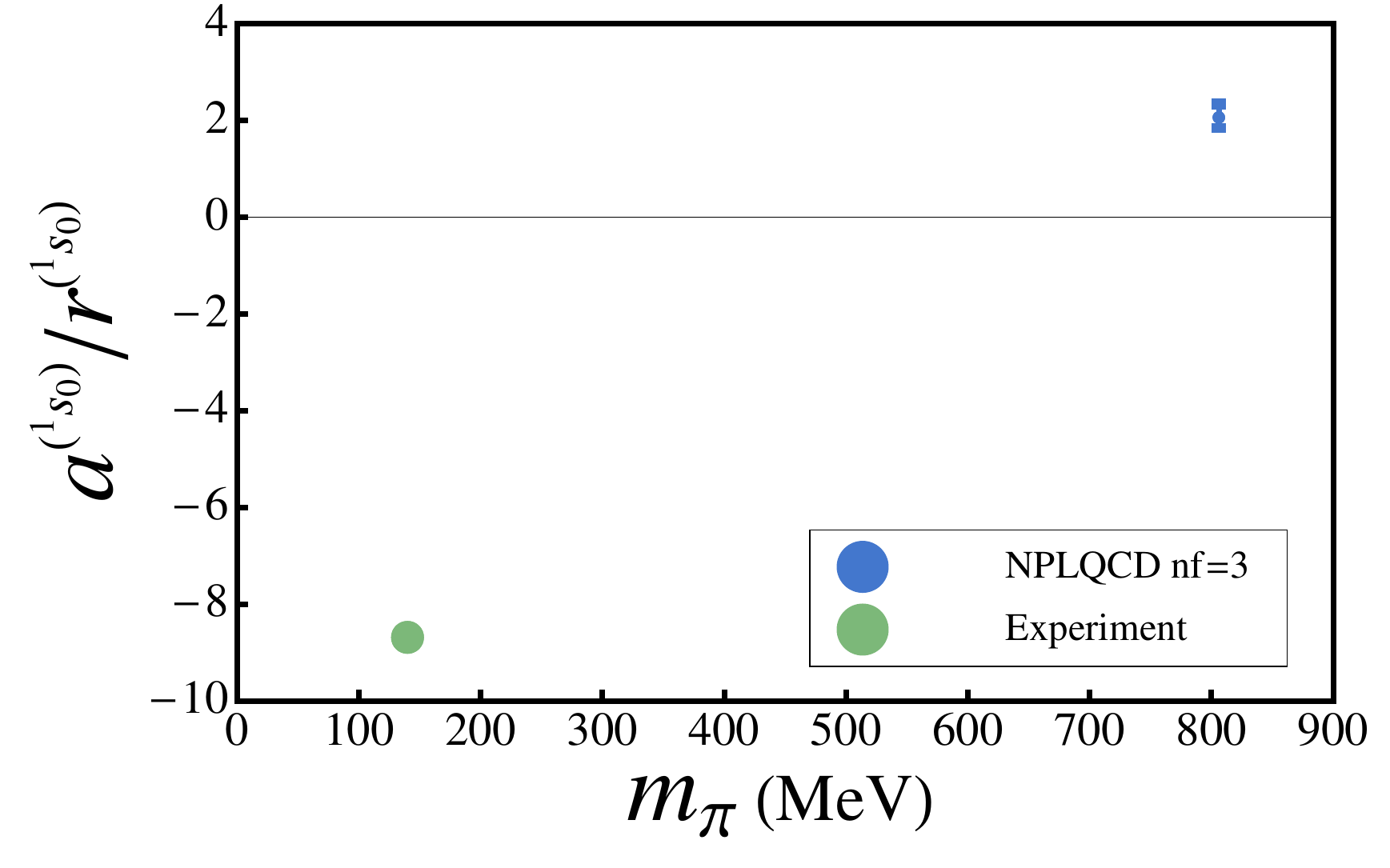}\ \ \
\includegraphics[width=3.2in,angle=0]{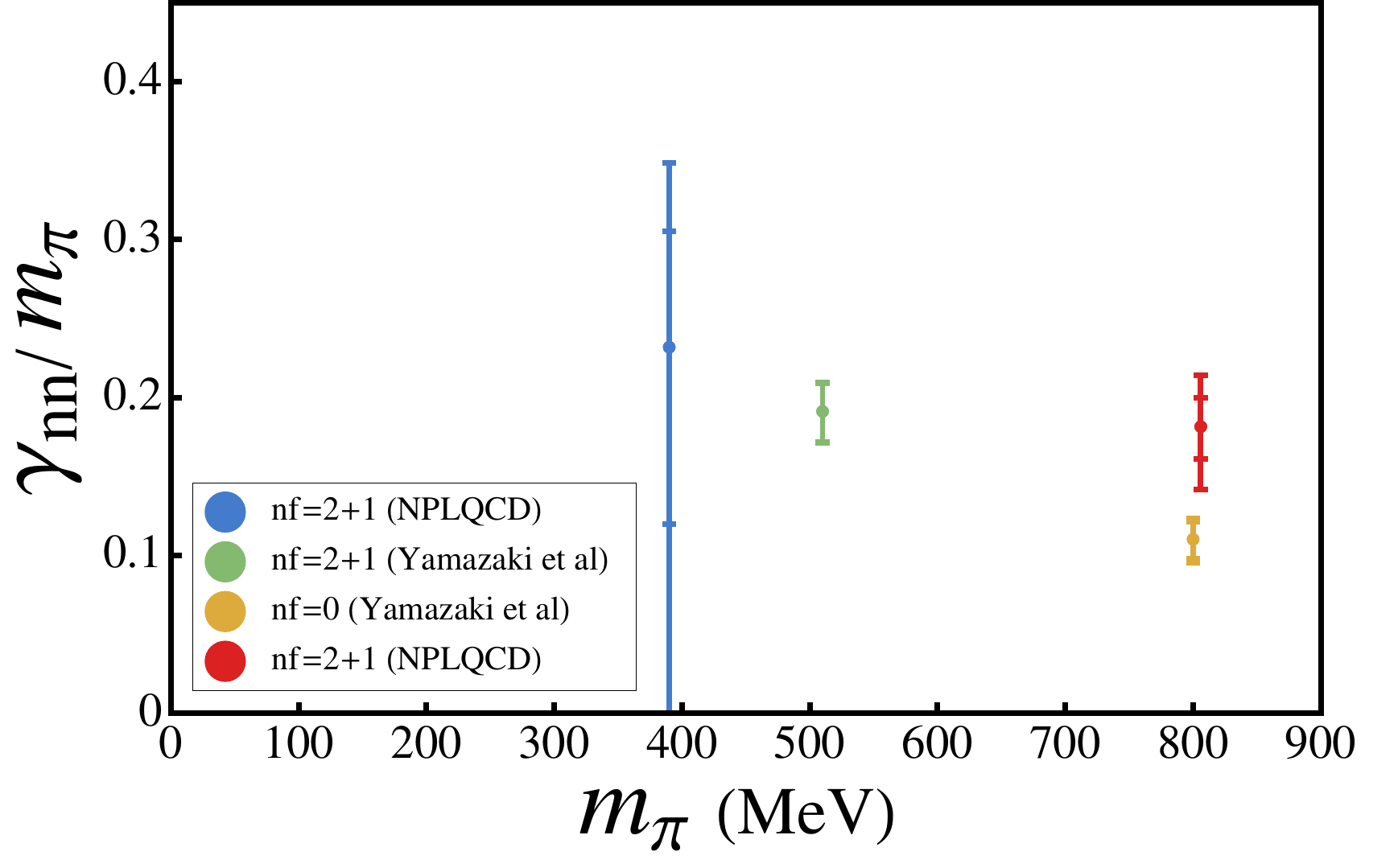}
}
\end{minipage}
\begin{minipage}[t]{16.5 cm}
\vspace{-0.1cm}
\caption{
The left panel shows the ratio of the scattering length to effective range in
the $\si$ channel.
The right panel shows the normalized di-neutron binding momentum 
versus the pion mass~\cite{Yamazaki:2011nd,Beane:2011iw,Beane:2012vq,Yamazaki:2012hi}. 
}
 \label{fig:nnar}
\end{minipage}
\end{center}
\end{figure}
The $\si$ channel is unnatural at the physical point with a very large
scattering length, but the system appears to be more natural at
heavier pion masses. Nonetheless, as shown in fig.~\ref{fig:nnar}
(left panel), the scattering length is approximately twice the
effective range at a pion mass of $m_\pi\sim \mpiMeV$, similar to the
$\siii$ channel.  In the right panel of fig.~\ref{fig:nnar}, the
di-neutron binding momentum $ \gamma_{nn}$ (related to the binding
energy by $B_{nn} = \gamma_{nn}^2/M_N$) normalized to the pion mass is
shown as a function of the pion mass. As in the $\siii$ channel, it
appears that the pion is not providing the only significant contribution to
the long-range component of the nuclear force. However, in contrast to
the $\siii$ channel, the $\si$-channel is clearly finely-tuned at the
physical light-quark masses.  The range of light-quark masses over which it is
fine-tuned requires further LQCD calculations to determine, and
eventual consideration of isospin violating effects due to light-quark
mass differences and electromagnetism. However, given the experimental
determinations of the $nn$, $np$ and $pp$ scattering lengths, these
effects are expected to be small.

It is interesting to note that the ratio of the scattering length to the
effective range in the two channels have very similar values at the
quark masses used in this work:
\begin{eqnarray}
a^{(\siii)}/r^{(\siii)} & = & 2.06^{+0.22}_{-0.18} {}^{+0.25}_{-0.19}
\ \ ,\ \ 
a^{(\si)}/r^{(\si)} \ = \ 2.02^{+0.23}_{-0.19} {}^{+0.29}_{-0.18}
\ \ \ ,
\label{eq:arcompare}
\end{eqnarray}
and that the scattering lengths in the two channels, and also the
effective ranges, are within $\sim 20\%$ of each other.  In the
large-N$_c$ limit of QCD, the nuclear forces in the two spin channels
are equal up to corrections suppressed by ${\cal
  O}(1/N_c^2)$~\cite{Kaplan:1995yg}, and the two channels transform in
the {\bf 6} of the Wigner SU(4) symmetry. 
In addition, inequalities for the binding energies of light nuclei in the
Wigner-symmetry limit have been found in Ref.~\cite{Chen:2004rq}.
 The closeness of the values
of the scattering parameters at $m_\pi\sim 800~{\rm MeV}$ is
consistent with the expectations of the large-N$_c$ limit of QCD.

\section{Conclusions and Discussions }
\label{sec:Conc}

\noindent We have presented the results of Lattice QCD calculations of
low-energy NN scattering phase-shifts and scattering parameters at the
SU(3) symmetric point with a pion mass of $m_\pi\sim 800~{\rm MeV}$.
For the first time, the effective ranges of the NN interactions have
been determined using lattice QCD. The calculated scattering lengths
and effective ranges indicate that the pion is not
the dominant contribution to the long range part of the nuclear force
at these large light-quark masses, as anticipated from the
single-hadron spectrum.  In both spin channels, the NN phase shifts 
change sign at higher momentum, near the start of the t-channel cut,
indicating that the nuclear interactions have a repulsive core even
for heavier quark masses.  This suggests that the form of the nuclear
interactions, and the effective potentials that will reproduce the
scattering amplitude below the inelastic threshold, is qualitatively
similar to the phenomenological potentials that describe the
experimental scattering data at the physical pion mass.

Both spin channels are, in a sense, more natural at $m_\pi\sim
800~{\rm MeV}$, where both satisfy $a/r\sim +2.0$, than at the physical
pion mass where $a^{(\si)}/r^{(\si)}\sim -8.7$ and
$a^{(\siii)}/r^{(\siii)}\sim +3.1$.  The relatively large size of the
deuteron compared with the range of the nuclear forces may persist
over a large range of light-quark masses, and therefore might, in
fact, not be usefully regarded as a fine-tuning in $n_f=2+1$ QCD, but
rather a generic feature.  The $\si$ channel, in contrast, is finely
tuned at the physical light-quark masses and it remains to be seen
over what range of masses this persists.

Our calculations were performed at a single pion mass with one lattice
spacing and in the absence of electromagnetic interactions. It should
be stressed that in the presence of fine-tuning, as in the $\si$
channel at the physical point, lattice-spacing artifacts can be
enhanced with respect to expectations based on naive
dimensional analysis and scaling arguments. In order to fully explore
the behavior of the scattering phase shifts and scattering parameters
with fully quantified uncertainties, along with the issues of
spin-flavor symmetry and fine tunings, calculations at multiple
lattice spacings and smaller light-quark masses are essential and are
planned for the future.

\vskip0.2in

\acknowledgments We thank R. Edwards and B. Jo\'{o} for help with
QDP++ and Chroma~\cite{Edwards:2004sx}.  We acknowledge computational
support from the USQCD SciDAC project, the National Energy Research
Scientific Computing Center (NERSC, Office of Science of the US DOE,
DE-AC02-05CH11231), the UW HYAK facility, LLNL, the PRACE Research
Infrastructure resource CURIE based in France at the Tr\`{e}s Grand
Centre de Calcul, TGCC, and the NSF through XSEDE resources under
grant number TG-MCA06N025.  SRB and PJ were partially supported by NSF
continuing grant PHY1206498. In addition, SRB gratefully acknowledges
the hospitality of HISKP and the Mercator programme of the Deutsche
Forschungsgemeinschaft.  The work of AP is supported by the contract
FIS2011-24154 from MEC (Spain) and FEDER.  H-WL and MJS were supported
in part by the DOE grant DE-FG03-97ER4014, and the NSF MRI grant
PHY-0922770 (HYAK).  KO was supported in part by DOE grants
DE-AC05-06OR23177 (JSA) and DE-FG02-04ER41302.  WD was supported by
the U.S. Department of Energy under cooperative research agreement
Contract Number DE-FG02-94ER40818 and by DOE Outstanding Junior
Investigator Award DE-SC000-1784.  The work of TL was performed under
the auspices of the U.S.~Department of Energy by LLNL under Contract
DE-AC52-07NA27344.  The work of AWL was supported in part by the
Director, Office of Energy Research, Office of High Energy and Nuclear
Physics, Divisions of Nuclear Physics, of the U.S. DOE under Contract
No. DE-AC02-05CH11231.  MJS thanks the Alexandar von Humboldt
foundation for the award that enabled his visit to the University of
Bonn, and the kind hospitality of Ulf Mei\ss ner and the
Helmhotz-Institut fur Strahlen- und Kernphysik at the University of
Bonn.


\end{document}